# Direct blockmodeling of valued and binary networks: a dichotomization-free approach

Carl Nordlund

carl.nordlund@liu.se

## Abstract

A long-standing open problem with blockmodeling is that it is explicitly intended for binary, not valued, networks. Whereas indirect methods exist that work reasonably well for partitioning valued networks, the underlying dilemma of comparing empirical valued blocks with ideal binary blocks remains. This is intrinsically problematic in the direct approach where partitions are solely determined through such comparisons. Addressing this, valued networks can either be dichotomized into binary versions or novel types of ideal valued blocks can be introduced. Both these workarounds are problematic in terms of interpretability, unwanted data reductions, and the often arbitrary setting of model parameters.

This paper proposes a dichotomization-free and parameter-free direct blockmodeling approach that effectively bypasses the dilemma with blockmodeling of valued networks. By generalizing the core-periphery heuristic of Borgatti and Everett (2000) and its extension by Nordlund (2018), this paper proposes an adaptive weighted correlation-based criteria function for direct blockmodeling. The proposed approach is directly applicable to both binary and valued networks, without any form of dichotomization or transformation of the valued (or binary) data at any point in the analysis, while still using the conventional set of ideal binary blocks from structural, regular and generalized blockmodeling.

Additionally, the proposed approach seemingly solves two other open problems with direct blockmodeling. First, its standardized goodness-of-fit measure allows for direct comparisons across solutions, within and between networks of different sizes, value types, and notions of equivalence. Secondly, through a well-known bias of point-biserial correlations, the approach puts a premium on optimal solutions that are closer to the mid-point density of the blockmodels. This, it is argued, translates into blockmodel solutions that are more intuitive and easier to interpret.

The approach is demonstrated by structural, regular and generalized blockmodeling applications of six classical binary and valued networks. Finding feasible and intuitive optimal solutions in both the binary and valued examples, the approach is proposed not only as a practical, dichotomization-free heuristic for blockmodeling of valued networks but also, through its additional benefits, as an alternative and would-be competitor to the conventional direct approach to structural, regular and generalized blockmodeling.

## Introduction

Blockmodeling[1] is concerned with the identification and classification of actors that are equivalent in some meaningful sense, and the analysis of the patterns of ties within and between such sets of equivalent actors. Given a network represented as an adjacency matrix, a blockmodel is created by sorting the rows and columns according to these sets of equvialent actors (also known as 'positions' or 'clusters'), where the delineated sub-matrix 'blocks'[2] within and between such positions are compared to different sets of ideal blocks. For an adequately fitted blockmodel, i.e. where the empirical blocks can be reasonable approximated in terms of such ideal blocks, the blockmodel can be reduced to a so-called blockimage, a representation of the original network that capture its underlying 'functional anatomy' (Nordlund, 2016, p. 161). Such blockimages can subsequently be compared across different networks, possibly of vastly different sizes, as well as interpreted and compared with the set of ideal blockimage templates that capture common archetypical structures, such as core-periphery structures, hierarchies, transitive structures, and cohesive subgroups (see Doreian, Batagelj, & Ferligoj, 2005, p. 236ff; Wasserman & Faust, 1994, p. 419,423; White, Boorman, & Breiger, 1976, pp. 742–744).

The notion of equivalence is foundational in blockmodeling, where different ideas and definitions of what is meant by actor equivalence determine the outcome and interpretation of a blockmodel analysis. The initial notion of equivalence was that of *structural* equivalence: formally introduced by Lorrain and White (1971), two actors are deemed structurally equivalent if they have identical ties to their common alters. Under this notion of equivalence, the empirical blocks in a structural equivalence blockmodel should ideally be either filled with ties – a so-called complete block – or void of any ties whatsoever – which constitute a null block. Loosening the criteria for structural equivalence, White and Reitz (1983, 1985; building on Sailer, 1978) proposed the notion of *regular* equivalence. Two actors are deemed regularly equivalent if they have ties to alters that in turn are also regularly equivalent. In regular equivalence blockmodeling, this kind of equivalence is captured by the regular block, an ideal block type where there is at least one tie in each row and column of the block. Supplementing the regular block, classical regular blockmodeling also includes the null block. Examples of complete, null and regular blocks are given in Figure 1 below, for both off-diagonal blocks (i.e. with ties between two sets of actors) and diagonal blocks (i.e. for ties among a single set of actors).

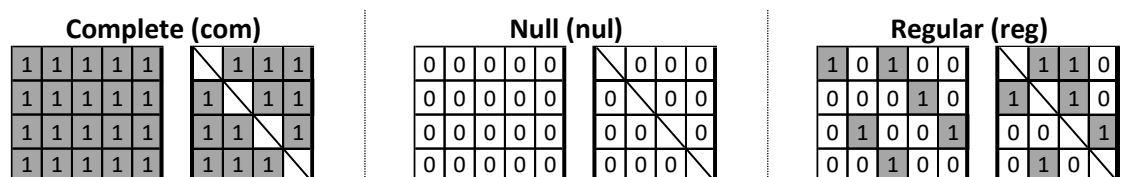

*Figure 1: Examples of complete, null and regular blocks located off or on the diagonal of a blockmodel*

Generalized blockmodeling is a broad analytical framework that allows for a wider variety of equivalences. Incorporating the three ideal blocks used to capture structural and regular equivalence, generalized blockmodeling also introduces additional types of ideal blocks (see Doreian et al., 2005, p. 212). Separating the criteria of regular blocks into rows and columns, respectively, the row-regular ideal block has at least one tie in each row of the block, and a column-regular has at least one tie in each block column. For row-functional blocks, there is exactly one tie in each row of the block (with corresponding column-wise criteria for column-functional blocks). A third novel type are the dominant-directional blocks, where the ideal row-dominant (column-dominant) block implies that at least one row (column) in the block is filled with ties. Examples of the directional varieties of the regular, functional and dominant block types are given in Figure 2 below, for off- and on-diagonal blockmodel locations.

---

[1] For a more detailed introduction to blockmodeling and role-analysis, see Wasserman and Faust (1994) and Doreian et al (2005). In addition to the deterministic kind of blockmodeling that this article is concerned with, there is also stochastic blockmodeling – see Wasserman and Faust (1994, Chapter 16), a chapter that also contains a section of goodness-of-fit measures of relevance for this current article.
[2] The term 'block' is here used to denote the specific submatrix in an adjacency matrix that is specified by one or two sets of actors.

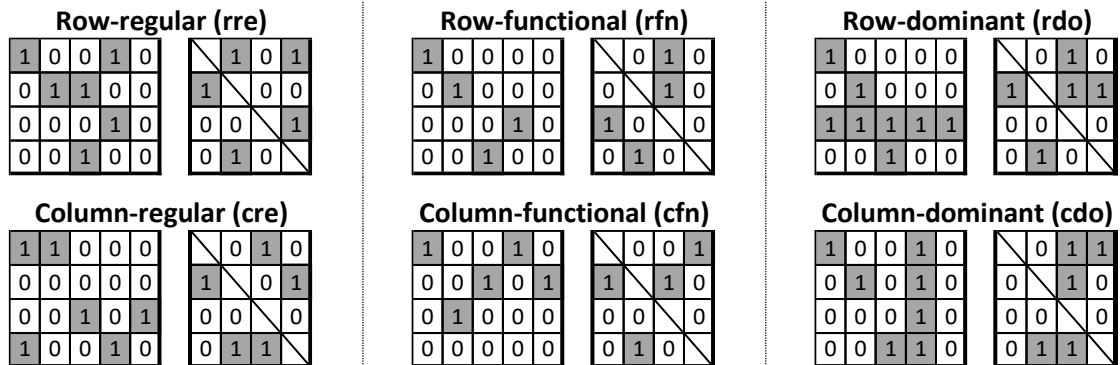

*Figure 2: Examples of ideal block types associated with generalized blockmodeling (see Nordlund 2016:162; Doreian et al 2004:212; Žiberna 2007a:108).*

The procedures to identify and classify actors into equivalence[3] sets can be divided into indirect and direct approaches. In the indirect approaches, pair-wise proxy measures for a given type of equivalence are first calculated[4] for each pair of actors in the network (see, e.g., Doreian et al., 2005, p. 177ff). Through a cluster analysis of these indirect measures of equivalence, identified classes of near-equivalent actors are subsequently used to sort the original adjacency matrix into a blockmodel. The emerging blocks are finally interpreted by comparing them with the set of ideal blocks of relevance for the specific notion of equivalence that the indirect measure is argued to capture. Contrasting this, the direct approach instead tries to fit a given set of ideal blocks directly with the original adjacency matrix. As an optimization problem where the number of positions are given a priori, a suitable relocation algorithm is used to find the particular actor partitions that minimizes the number of inconsistencies between the observed empirical blocks and a given set of ideal blocks that represents a specific type of equivalence. Although computationally heavy for larger networks and with the risk of search algorithms possibly failing to find optimal solutions, direct approaches nevertheless have several advantages over the indirect approaches (see Doreian et al., 2005, p. 184; Doreian, 2006, p. 124). The most profound and, in this case, crucial advantage of direct approaches is the lack of indirect measures or algorithms that can capture the more complex and varied ideal block types as found in generalized blockmodeling. In what follows, the focus is thus exclusively on direct blockmodeling approaches.

---

[3] Or, more commonly, near-equivalent. It is rare that actors in real-world empirical networks are perfectly equivalent, meaning that classes/subsets of actors typically contain some variance with respect to how structural, regular or otherwise equivalent they are.

[4] The classical indirect measure of structural equivalence of two actors involves the comparison of their corresponding row- and column vectors in the adjacency matrix, often using either Pearson correlations, Euclidean distances or Hamming distances to compare these vectors. For regular equivalence, White and Reitz (White & Reitz, 1983, 1985; also see varieties by Žiberna, 2007) introduced the REGE algorithm, a point-scoring algorithm that, it is argued, produces adequate pair-wise measures of regular equivalence of actors.

### The dilemma with blockmodeling of valued networks

An open and seemingly sticky problem in blockmodeling is that it is explicitly geared at binary, rather than valued, networks (Doreian, 2006; Nordlund, 2007, 2016; Žiberna, 2007b). The underlying dilemma is that the standard set of ideal blocks (Figure 1 and Figure 2) are all specified in terms of binary tie patterns, making it a non-trivial exercise to match and count inconsistencies between such ideal binary ties and empirical valued ties. This is of particular concern in the direct approaches where the identification and clustering of equivalent actors stem directly from such comparisons, but it is equally problematic when interpreting blocks for partitions derived from indirect methods or for given, prespecified hypothetical partitions.

Addressing this dilemma, a number of workarounds have been proposed. The classical yet obviously crude approach is to simply transform the network into a binary version through dichotomization. Instead of picking a single network-wide dichotomization threshold on statistical, theoretical or arbitrary grounds, an alternative is to instead apply a series of threshold values at periodic intervals, repeating and comparing the blockmodeling analyses for a larger ensemble of dichotomized binary networks (e.g. Doreian, 1969).

The opposite type of workaround is to instead introduce novel types of ideal blocks that cater for valued ties, such as the sum- and max-regular blocks proposed by Žiberna (2007a). In these novel ideal blocks, designed explicitly for valued networks, a threshold parameter is nevertheless needed to distinguish between the fulfillment or not of the specific block criteria. In a similar vein, Nordlund (2007) suggests a novel heuristic for the identification of regular blocks in valued networks, a heuristic that equally requires the choosing of a suitable threshold to distinguish between prominent and non-prominent ties.

### A novel direct approach to blockmodeling of binary and valued networks

This paper proposes a direct blockmodeling approach that effectively bypasses the dilemma with blockmodeling of valued networks. Inspired by the core-periphery heuristic of Borgatti and Everett (2000), the herein proposed approach replaces the classical inconsistency-counting criteria functions in direct blockmodeling with a generalized version of the weighted correlation coefficient approach suggested in Nordlund (2018). Applicable in both exploratory and confirmatory studies of both structural, regular and generalized blockmodeling, possibly with prespecified partitions, blockimages, or both, it is argued that the proposed approach not only represents a practical dichotomization-free solution to the valued blockmodeling problem but also provides several additional advantages compared to existing direct blockmodeling approaches, for binary and valued networks alike.

First, the proposed approach is equally applicable to binary as well as valued networks, without any type of dichotomization, transformation or reduction of the data resolution of the valued network data at any stage in the analysis. Despite this, secondly, the proposed approach still uses the conventional set of binary ideal blocks of structural, regular and generalized blockmodeling, without modifying or introducing any specialized valued versions of ideal blocks (cf. Žiberna, 2007a). In line with the generalized blockmodeling framework (Doreian et al., 2005, p. 245ff), the proposed approach also allows for the introduction of additional ideal blocks, specified in terms of binary tie patterns, including 'overlayed' blocks like the combined dependency-dominance blocks as suggested in Nordlund (2018).

With the weighted correlation coefficient as the criteria function, the relative contribution of each empirical block to the final goodness-of-fit measure can be kept proportional to the relative block sizes of a blockmodel, irrespective of the type of ideal block that is tested for. For regular blocks, this bypasses the need to have different ways to measure block inconsistencies pending on whether we are conducting a classical regular equivalence blockmodeling or working with the broader range of blocks in generalized blockmodeling (see Doreian et al., 2005, pp. 188, 224). By following this principle of size-proportional influence to the overall goodness-of-fit measure, the need to balance the penalties for different types of ideal blocks is greatly diminished.

Whereas "the magnitudes of the [conventional inconsistency-counting] criterion functions cannot be compared across different types of blockmodels" (Doreian, 2006, p. 127), it is here argued that a standardized correlation-based measure makes such comparisons possible. The herein proposed approach thus also provides a solution to this hitherto unsolved problem of comparing goodness-of-fit measures across different blockmodel solutions, with possibly vastly different network sizes and types of equivalences, for both binary and valued networks.

Two additional advantages stem from using a correlation-based goodness-of-fit function for direct blockmodeling. The first is the inherent bias of point-biserial correlations (i.e. a correlation when a naturally dichotomous variable is correlated with a continuous variable) that favors solutions where a split of the data along the dichotomous variable results in equally large subsets. Although this is occasionally perceived as a problem in various applications, it is here argued[5] to be an advantage in this context, as it puts a premium on ideal blockmodels that are closer to the mid-point (0.5) density than further away from it. Stemming from this, the second advantage with using a correlation-based heuristic for direct blockmodeling is that it is less likely to find multiple, equally optimal solutions than what is the case for the conventional inconsistency-minimizing approach. Although this makes it imperative to also explore near-optimal solution, possibly classifying near-optimal solutions as equally optimal as the max-correlation solution, the correlation-based measure can act as a guide in choosing which out of many near-optimal solutions to use as a representative blockmodel.

The final and possibly crucial advantage with the herein proposed approach is that for the binary networks so far tested, it seem to capture either identical or very similar solutions as those found when using the existing conventional approach for direct binary blockmodeling. As there are other advantages of the herein proposed approach, even when we stay within the realm of binary empirical networks, the weighted correlation coefficient heuristic in this paper is therefore not only proposed as a premier solution to direct blockmodeling of valued networks, but also as an alternative to the conventional direct approach to binary network blockmodeling.

The remainder of this article is divided into three parts. Starting off with the Borgatti-Everett core-periphery heuristic and its power-relational extensions proposed by Nordlund (2018), the next section will discuss the foundations for the proposed heuristic, providing pseudocode for how various ideal binary blocks can be captured. This is followed by an example section. Beginning with a trivial toy network, two classical binary networks will be analyzed, followed by three valued networks: the Hlebec notesharing data (Hlebec, 1996; Žiberna, 2007a), the Ucinet river primates network (Borgatti & Everett, 2000, p. 380), and the Freeman's EIES friendship networks (Freeman & Freeman, 1980). The paper is concluded with a short summary of the proposed heuristics, observations from the provided examples, and potential shortcomings.

## Specification of proposed heuristic

The direct approach for binary and valued generalized blockmodeling proposed in this paper takes inspiration from the core-periphery measure and heuristic proposed by Borgatti and Everett (2000) and the extensions to this approach as proposed by Nordlund (2018). Starting off with these correlation-based approaches for identifying core-periphery structures, this section describes how these can be generalized and extended. This is followed by specifications on how the ideal blocks in structural, regular and generalized[6] equivalence can be identified, measured and combined into the aggregate goodness-of-fit measure proposed in this paper.

---

[5] This will be discussed and detailed in the next section of the paper.

### The Borgatti-Everett measure of core-periphery structures

To capture the classical categorical core-periphery notion of dense cores and internally disconnected peripheries for a hypothetical 2-positional blockmodel, Borgatti and Everett (2000) suggest using the (population-wise) correlation coefficient resulting from correlating all intra-core cells unity and corresponding intra-peripheral cells with zero. This measure thus reaches unity when the set of core actors form a clique[7] and when there are no ties between the remaining set of peripheral actors. Borgatti and Everett demonstrate how this correlation can be used as a goodness-of-fit measure in a relocation algorithm for finding optimal core-periphery partitions. Exemplifying the measure using the hypothetically partitioned 6-actor directional binary network in Figure 3 below, the calculation of the core-periphery correlation for this particular blockmodel is given in Table 1.

|   |   | C |   |   | P |   |   |
|---|---|---|---|---|---|---|---|
|   |   | a | b | c | d | e | f |
| C | a |   | 1 | 1 | *0* | *1* | *1* |
|   | b | 0 |   | 1 | *0* | *0* | *0* |
|   | c | 1 | 1 |   | *0* | *0* | *1* |
| P | d | *0* | *1* | *1* |   | 0 | 0 |
|   | e | *1* | *0* | *0* | 0 |   | 0 |
|   | f | *0* | *0* | *1* | 0 | 1 |   |

*Figure 3: 6-actor directional binary network example for core-periphery correlation testing*

| Block | C→C (Complete) | | | | | | P→P (Null) | | | | | |
|---|---|---|---|---|---|---|---|---|---|---|---|---|
| Cell | a→b | a→c | b→a | b→c | c→a | c→b | d→e | d→f | e→d | e→f | f→d | f→e |
| Observed (X) | 1 | 1 | 0 | 1 | 1 | 1 | 0 | 0 | 0 | 0 | 0 | 1 |
| Ideal (Y) | 1 | 1 | 1 | 1 | 1 | 1 | 0 | 0 | 0 | 0 | 0 | 0 |

Nbr of inconsistencies:   2                    corr(X,Y):   2/3

*Table 1: Calculating the Borgatti-Everett core-periphery measure for the blockmodel in Figure 3*

By fitting an empirical network to the prespecified complete and null blocks, the Borgatti-Everett heuristic for identifying core-periphery structures is a direct approach for structural equivalence blockmodeling. What differs from conventional direct blockmodeling is its criteria function: the classical procedure of counting and minimizing the total number of inconsistencies (i.e. the Hamming distance between observed and ideal value-pairs) is in the Borgatti-Everett approach replaced by a correlation[8] measure. As exemplified in Table 1, the complete and null block tests are each represented by subsets of value-pairs capturing the empirical and ideal values for each cell, which when combined form the whole set of observed and ideal ties for the overall measure of fit.

---

[6] Of the six types of ideal blocks typically specified for generalized blockmodeling – the row- and column-varieties of, respectively, regular, functional and dominant ideal blocks (see Figure 2 above) – this paper will specify and provide pseudocode for the regular and functional types, leaving the row- and column-dominant ideal blocks for would-be future implementations.

[7] I.e. where all possible ties exist and where they have equal tie values (i.e. unity in binary networks).

[8] A correlation coefficient of two binary variables is also called the phi ($\varphi$) coeffcient or, as in Hubert and Baker (1978), the $\Gamma$ index. As the correlation coefficient is "designed" to measure the linear association between two continuous variables, its applicability for dichotomous data has been criticized (see also Carrington, Heil, & Berkowitz, 1979, p. 223ff). Borgatti and Everett however point out that they are merely using the Pearson correlation coefficient as a measure of association between two variables, without the associated inferential test (2000, p. 379; cf. Arabie, Boorman, & Levitt, 1978). This is exactly the setting in which this current paper and its proposed heuristic operate.

While the ideal ties for a core-periphery structure indeed are binary (*Y* in Table 1), the observed values (*X* in Table 1) could be either binary or valued (Borgatti & Everett, 2000, p. 379). This heuristic is thus applicable to both binary and valued network, where the full range of would-be tie values are utilized in the fitting and measuring of both optimal and hypothetical core-periphery structures.

As the values in *Y* are always binary, this correlation is also known as the point-biserial correlation, expressed in the following formula:

$$r_{pb} = \frac{M_1 - M_0}{s_n} \sqrt{\frac{n_1 n_0}{n^2}} \qquad \text{(Eq. 1)}$$

where $M_1$ and $M_0$ are the mean values for the subsets of *X* values with corresponding *Y* values being, respectively, 1 and 0, $s_n$ being the population-wide standard deviation of all *X* values, $n_1$ and $n_0$ being the number of *Y* values that are, respectively, 1 and 0, and *n* being the total length of the list *L*.

The point-biserial representation (Eq. 1) is particularly instructive for understanding an important aspect of its usage as a blockmodeling criteria function, such as the Borgatti-Everett core-periphery measure. Dividing up the values in the *X* vector into two sets, one for respective *Y* value, the left-side factor of the point-biserial formula is concerned with the differences in mean values between these sets. The right-hand factor in the formula, however, is only concerned with the relative sizes of these two sets, a factor that reaches a maximum of 0.5 when $n_1$ equals $n_0$. As a measure of goodness-of-fit in the Borgatti-Everett core-periphery heuristic and, as elaborated below, its generalization as a criteria function for direct structural equivalence, this preference for an equal number of 1:s and 0:s in the *Y* vector translates into a bias towards optimal blockmodel solutions whose ideal counterparts are closer to a mid-way density. Whereas this particular 50/50 phenomena of point-biserial correlations has been deemed problematic and worthy of 'corrections' in certain context (see, e.g., Kemery, Dunlap, & Griffeth, 1988; cf. Williams, 1990), it is here argued to instead be beneficial in the blockmodeling context by its inherent 'bias', or rather preference, for more interpretable solutions.

The first step towards the approach suggested in this paper is a generalization of the Borgatti-Everett heuristic for identifying core-periphery structures. By allowing for arbitrary configurations of complete and null blocks, for two or more sets of equivalent actors, we obtain a direct approach for structural equivalence blockmodeling that, similar to the original core-periphery heuristic, is equally applicable to binary and valued data.

Whereas the standard (point-biserial) correlation coefficient works well for capturing the complete and null ideal block types, it is less useful for capturing the more complex tie patterns of the ideal blocks used in regular and generalized blockmodeling. Rather, building on and generalizing how the complex ideal block types of dependency and dominance were operationalized in Nordlund (2018), the weighted correlation coefficient is instead used in the specification which we now turn to.

### A weighted correlation-based direct blockmodeling approach[9]

Given the adjacency matrix *A* of a (possibly directional) network, where $a_{row,col}$ is the tie value from actor *row* to actor *col* (with self-ties undefined), the set of actors are partitioned into *k* non-overlapping subsets $P=\{P_1,..,P_k\}$. The blockmodel is given by *B(A,P)*, where $B_{r,c}$ corresponds to the particular empirical block of *A* that is delineated by the actors in $P_r$ and $P_c$. For diagonal blocks, *r* equals *c*. We also have a specified ideal blockimage *I*, where $I_{r,c}$ represents the ideal block to be compared with the relational patterns of block $B_{r,c}$. For structural equivalence blockmodeling, *I* contains either complete or null blocks.

---

[9] A demonstrational Windows client that implements the proposed heuristic is available at http://carlnordlund.net

To calculate the weighted correlation measure for a specific partition, we compile a list *L* of ordered value-triplets (*x*; *y*; *w*), values that correspond to, respectively, an observed value (*x*), an ideal value (*y*), and a weight (*w*) for each (*x*; *y*) value-pair. Given the blockmodel *B* and blockimage *I*, the subset of value-triplets for the specific empirical and ideal block at row *r* and column *c* of the blockmodel is denoted by $f_{BLK}$ (*A*, $P_r$, $P_c$) where BLK is the specific ideal block – $I_{r,c}$ – that is tested for. Concatenating the value-triplets for all blocks into the list *L*, the goodness-of-fit for this blockmodel and partition is the weighted correlation coefficient $r_w$ where each value-pair in the combined *X* and *Y* vectors uses the corresponding weights in the combined *W* vector:

$$r_w = \frac{\sum w_i(x_i - \bar{x})(y_i - \bar{y})}{\sqrt{\sum w_i(x_i - \bar{x})^2}\sqrt{\sum w_i(y_i - \bar{y})^2}} \quad \text{(Eq. 2)}$$

where $x_i$ and $y_i$ are, respectively, the observed and ideal values for value-triplet *i*, $w_i$ is the weight for the ($x_i$; $y_i$) pair, and where the weighted means are $\bar{x} = \sum w_i x_i / \sum w_i$ and $\bar{y} = \sum w_i y_i / \sum w_i$.

Given the empirical block $B_{i,j}$ as specified by the adjacency matrix *A* and the two subsets of actors $P_i$ and $P_j$, the pseudocode for generating the (*x*; *y*; *w*) value-triplets to test this empirical block with the complete and null blocks, respectively, is given in Table 2 below. As the criteria for these blocks imply that each cell is checked exactly once, the weighting for each value-pair is equal to unity for these blocks.

| $f_{COMPLETE}$(A, $P_i$, $P_j$): | $f_{NULL}$(A, $P_i$, $P_j$): |
|---|---|
| $L_{i,j}$={ } <br> foreach r in $P_i$: <br>   foreach c in $P_j$: <br>     if r<>c: <br>       $L_{i,j}$ ←($a_{r,c}$; 1; 1) <br> return $L_{i,j}$ | $L_{i,j}$={ } <br> foreach r in $P_i$: <br>   foreach c in $P_j$: <br>     if r<>c: <br>       $L_{i,j}$ ←($a_{r,c}$; 0; 1) <br> return $L_{i,j}$ |

*Table 2: Pseudocode to generate correlation value-triplets for complete and null blocks[10]*

The regular block criteria stipulates that there must be *at least* one tie in each row and column, respectively, of an empirical block. It is thus only necessary to check the largest tie values[11] in each row and column, respectively. For the correlation-based heuristic here proposed, this would however imply that the number of observed-vs-ideal value-pairs for checking a regular block would be far fewer than the number of value-pairs needed to check for a complete or a null block. For instance, for an off-diagonal block with 4 rows and 6 columns, only 10 observations would be needed to check for a regular block, whereas a null or a complete block would need to observe all 24 cells in this particular block. Thus, to ensure that each block contributes in a size-proportional manner to the overall goodness-of-fit measure for the blockmodel as a whole, the weights for the checked empirical-ideal value-pairs are adjusted accordingly. In the example with a 4-by-6 block, each of the 10 value-pairs would then carry a weight of 2.4, adjusting so that the relative contribution for this block to the aggregate correlation would be proportional to its size. Pseudocode to generate value-triples for regular blocks in this way is given in Table 3 below.

---

[10] It is here assumed that the network contains no self-ties. If self-ties exist and are allowed, the if-statements have to be removed.

[11] In binary networks, this would then ideally be unity.

```
f_REGULAR(A, P_i, P_j):
L_{i,j}={ }
w=(|P_i|·(|P_j| - [i=j])) / (|P_i| + |P_j|)
foreach r in P_i:
   L_{i,j} ←(max_c a_{r,c}; 1; w)
foreach c in P_j:
   L_{i,j} ←(max_r a_{r,c}; 1; w)
return L_{i,j}
```

Table 3: Pseudocode to generate correlation value-triplets for regular blocks (Iverson square brackets, i.e. true→1, false→0; $|P_i|$: cardinal number of actors in position $P_i$; $max_x\, a_{x,y}$: the maximum value in column y of the A matrix)

Separating the criteria for regular blocks into rows and columns, respectively, subsequently also adjusting the weights for each checked value-pair, pseudocode for row-regular and column-regular blocks are found in Table 4 below. For row-regular blocks, the maximum value in each block will be correlated with unity, weighting this by the number of columns in the block (adjusted for would-be self-ties). For column-regular blocks, a corresponding weighting is done with respect to the number of block rows.

| f_ROW-REGULAR(A, P_i, P_j): | f_COLUMN-REGULAR(A, P_i, P_j): |
|---|---|
| L_{i,j}={ }<br>w=\|P_j\| - [i=j]<br>foreach r in P_i:<br>   L_{i,j} ←(max_c a_{r,c}; 1; w)<br>return L_{i,j} | L_{i,j}={ }<br>w=\|P_i\| - [i=j]<br>foreach c in P_j:<br>   L_{i,j} ←(max_r a_{r,c}; 1; w)<br>return L_{i,j} |

Table 4: Preudocode to generate correlation value-triplets for row-regular (left) and column-regular (right) blocks ($max_c\, a_{r,c}$: the maximum value in row vector r in A; other notation as in Table 3)

For the row-functional ideal block, there is *exactly one* tie on each row in the block. In the conventional generalized blockmodeling framework for binary networks (see Doreian et al., 2005, p. 224), the incurred penalty for a block row depends on whether it is empty or contains more than one tie. For an empty block row, the incurred penalty is equal to the length of this row, i.e. the number of columns that the block has. Conversely, a singular penalty is incurred for the number of ties in a block row exceeding one. Corresponding penalties are incurred for column-functional blocks.

Translating the criteria for row- and column-functional blocks into value-triplets for the correlation-based approach is less straight-forward than what was the case for the previously specified ideal blocks. In Nordlund (2018, p. 354ff), the suggested approach for capturing row-functionality is to correlate the largest value in a block row with unity and all remaining values with zero. For block rows completely devoid of ties, only one of these zero values would then be correlated with unity where remaining zeros were correlated with zero. Inspired by the full-row penalty for empty block rows in the conventional generalized approach, we instead opt for the more "penalizing" alternative suggested in Nordlund (2018, p. 355, note 11) where a similar full-row weighted penalty is incurred for would-be empty block rows. Pseudocode for capturing row- and column-functional blocks in this way is given in Table 5 below.

| $f_{\text{ROW-FUNCTIONAL}}(A, P_i, P_j)$: | $f_{\text{COLUMN-FUNCTIONAL}}(A, P_i, P_j)$: |
|---|---|
| ```
L_{i,j}={ }
w=|P_j| - [i=j]
foreach r in P_i:
  max_r = max_c a_{r,c}
  if max_r = 0:
    L_{i,j} ←(0; 1; w)
  else:
    index = -1
    foreach c in P_j:
      if a_{r,c} = max_r:
        index_c = c
    foreach c in P_j:
      if c = index_c:
        L_{i,j} ←(a_{r,c}; 1; 1)
      else:
        L_{i,j} ←(a_{r,c}; 0; 1)
return L_{i,j}
``` | ```
L_{i,j}={ }
w=|P_i| - [i=j]
foreach c in P_j:
  max_c = max_r a_{r,c}
  if max_c = 0:
    L_{i,j} ←(0; 1; w)
  else:
    index = -1
    foreach r in P_i:
      if a_{r,c} = max_c:
        index_r = r
    foreach r in P_i:
      if r = index_r:
        L_{i,j} ←(a_{r,c}; 1; 1)
      else:
        L_{i,j} ←(a_{r,c}; 0; 1)
return L_{i,j}
``` |

*Table 5: Preudocode to generate correlation value-triplets for row-functional (left) and column-functional (right) blocks (notation as in Table 3 and Table 4 )*

## Example section

This section will demonstrate the proposed direct blockmodeling approach on three binary and three valued networks. Starting off with a trivial 10-actor toy network previously used to demonstrate core-periphery structures (Borgatti & Everett, 2000; Nordlund, 2018), this is followed by analyses of two classical binary networks previously used to demonstrate generalized blockmodeling: friendship among the Transatlantic industries Little League baseball team (Fine, 1987; Doreian et al., 2005, p. 48), and the Kansas Search and Rescue data (Drabek, Tamminga, Kilijanek, & Adams, 1981; Doreian et al., 2005, p. 203). For these binary networks, the results from the proposed correlation-based approach will be compared with corresponding results from using the conventional direct approach for structural, regular and generalized blockmodeling.

Following the binary examples are three valued examples: the Hlebec notesharing network (Hlebec, 1996; Žiberna, 2007a), the co-presence of Florida primates (Borgatti & Everett, 2000, p. 380), and the Freeman EIES friendship data (Freeman & Freeman, 1980). These examples will demonstrate both how the proposed goodness-of-fit measure can be used for finding optimal partitions, blockimages, or both, as well as for hypothesis testing with given partitions.

### Binary example 1: The BEfig1 core-periphery network

The first example is the 'BEfig1' network in Figure 4 below (Borgatti & Everett, 2000, p. 377; Nordlund, 2018, p. 358), a trivial binary, symmetric 10-actor network introduced by Borgatti and Everett to exemplify how a typical core-periphery structure could look like. With the default Borgatti-Everett core-periphery specification, i.e. prespecifying the blockimage with a complete block for intra-core ties, a null block for intra-peripheral ties, and a 'do-not-care' block for core-periphery ties, both the correlation-based and inconsistency-based approaches find a singular ideal solution (*corr*=1.000; *penalty*=0) when actors 1 to 4 constitute the core, i.e. the outlined blockmodel in Figure 4 below.



|   | 1 | 2 | 3 | 4 | 5 | 6 | 7 | 8 | 9 | 10 |
|---|---|---|---|---|---|---|---|---|---|---|
| 1 | \ | 1 | 1 | 1 | 1 | 0 | 0 | 0 | 0 | 0 |
| 2 | 1 | \ | 1 | 1 | 0 | 1 | 1 | 1 | 0 | 0 |
| 3 | 1 | 1 | \ | 1 | 0 | 0 | 0 | 1 | 1 | 0 |
| 4 | 1 | 1 | 1 | \ | 1 | 0 | 0 | 0 | 0 | 1 |
| 5 | 1 | 0 | 0 | 1 | \ | 0 | 0 | 0 | 0 | 0 |
| 6 | 0 | 1 | 0 | 0 | 0 | \ | 0 | 0 | 0 | 0 |
| 7 | 0 | 1 | 0 | 0 | 0 | 0 | \ | 0 | 0 | 0 |
| 8 | 0 | 1 | 1 | 0 | 0 | 0 | 0 | \ | 0 | 0 |
| 9 | 0 | 0 | 1 | 0 | 0 | 0 | 0 | 0 | \ | 0 |
| 10 | 0 | 0 | 0 | 1 | 0 | 0 | 0 | 0 | 0 | \ |

*Figure 4: The 'BEfig1' core-periphery example, blocked according to the intuitive core (see Borgatti & Everett, 2000, p. 377; Nordlund, 2018, p. 358)*

Allowing for any non-trivial 2-positional configuration of complete and null blocks, i.e. a direct exploratory structural equivalence blockmodel analysis of the BEfig1 network, both the penalty-based and the correlation-based heuristics confirm that a default core-periphery structure with null blocks in the off-diagonal blocks constitutes the optimal solution (*corr*=0.5837; *pen*=16) – see Table 6 below. However, whereas both heuristics place actors 1 to 4 in that solution, the penalty-based heuristics finds two additional, equally optimal, solutions that also place actor 5 and 8, respectively, in the core. Even though these solutions indeed contain 16 inconsistencies, the correlation-based approach deems these additional 5-actor-core solutions as slightly less optimal (*corr*=0.5645) than the more intuitive 4-actor core (i.e. *corr*=0.5837).

**Correlation-based optimal solution [com, nul], k=2**

| # solutions | Blockimage[12] | corr (penalties) | P1 | P2 |
|---|---|---|---|---|
| 1 | [com,nul],[nul,nul] | 0.5837 (16) | 1,2,3,4 | rest |
| 1 | [com,com],[com,nul] | 0.5324 (24) | 2,3,4 | rest |

**Inconsistency-based optimal solutions [com, nul], k=2**

| # solutions | Blockimage | Penalties (corr) | P1 | P2 |
|---|---|---|---|---|
| 3 | [com,nul],[nul,nul] | 16 (0.5837) | 1,2,3,4 | rest |
|   | -"- | 16 (0.5645) | 1,2,3,4,5 | rest |
|   | -"- | 16 (0.5645) | 1,2,3,4,8 | rest |
| 3 | [com,com],[com,nul] | 22 (0.4664) | 2,3 | rest |
|   | -"- | 22 (0.4664) | 2,4 | rest |
|   | [–,com],[com,nul] | 22 (0.3840) | 2 | rest |

*Table 6: Optimal solutions for 2-positional structural equivalence blockmodeling of the 'BEfig1' network*

The second-most optimal 2-positional blockimage is the default core-periphery structure with complete blocks in the off-diagonal blocks – see Table 6 above. However, whereas the correlation-based approach finds a singular optimal solution when the core consists of actors 2, 3 and 4 (*corr*=0.5324; *pen*=24), the conventional direct approach once again provides three equally optimal partitions (*pen*=22), none of which are the same as the 2,3,4-core partition suggested by the correlation-based approach. Whereas two of these optimal solutions have a corresponding correlation of 0.4664, the solution with the singleton actor in the core only has a corresponding correlation-based goodness-of-fit measure of 0.3840.

Repeating the free-search optimization with three types of ideal blocks – complete, regular, and null blocks (in this order of priority) – both heuristics find identical ideal solutions as given in Table 7. In

---
[12] In the notation used here, $[a_{11},a_{12}],[a_{21},a_{22}]$ indicates a 2-by-2 blockimage where block position $a_{rc}$ is the ideal block between actor set $P_r$ and $P_c$. Dashes (–) are do-not-care or not-applicable, e.g. for singleton positions.



addition to the intuitive clique core with actors 1 to 4, there is also an alternative ideal solution where actors 2 to 5 form a core in the form of a regular block.

**Optimal solutions (both heuristics) [com, reg, nul], k=2**

| # solutions | Blockimage | corr / Penalty | P1 | P2 |
|---|---|---|---|---|
| 2 | [com,reg],[reg,nul] | 1.0000 / 0 | 1,2,3,4 | rest |
|   | [reg,reg],[reg,nul] | 1.0000 / 0 | 2,3,4,5 | rest |

*Table 7: Optimal solutions for 2-positional generalized equivalence blockmodeling (complete, regular and null ideal blocks; this order of priority) of the 'BEfig1' network*

## Binary example 2: Little League baseball teams (Transatlantic Industries)

From data collected by Fine (1987) on friendship nominations within Little League baseball teams, Doreian and colleagues look closer at one of these teams – the Transatlantic Industries team – to demonstrate aspects of direct structural equivalence blockmodeling and the enhanced features of the generalized blockmodeling framework they propose. With each of the 13 team members nominating three[13] friends, the adjacency matrix for this binary, directed Transatlantic Industries network is given in Table 8 below.

|           | Ron_1 | Tom_2 | Frank_3 | Boyd_4 | Tim_5 | John_6 | Jeff_7 | Jay_8 | Sandy_9 | Jerry_10 | Darrin_11 | Ben_12 | Arnie_13 |
|---|---|---|---|---|---|---|---|---|---|---|---|---|---|
| Ron_1     |   | 0 | 1 | 1 | 1 | 0 | 0 | 0 | 0 | 0 | 0 | 0 | 0 |
| Tom_2     | 1 |   | 1 | 0 | 0 | 0 | 0 | 0 | 0 | 0 | 1 | 0 | 0 |
| Frank_3   | 1 | 0 |   | 1 | 0 | 0 | 0 | 0 | 0 | 0 | 1 | 0 | 0 |
| Boyd_4    | 1 | 1 | 1 |   | 0 | 0 | 0 | 0 | 0 | 0 | 0 | 0 | 0 |
| Tim_5     | 1 | 0 | 1 | 1 |   | 0 | 0 | 0 | 0 | 0 | 0 | 0 | 0 |
| John_6    | 0 | 0 | 0 | 0 | 1 |   | 0 | 0 | 0 | 0 | 0 | 1 | 1 |
| Jeff_7    | 0 | 1 | 0 | 0 | 0 | 0 |   | 1 | 1 | 0 | 0 | 0 | 0 |
| Jay_8     | 0 | 1 | 0 | 0 | 0 | 0 | 1 |   | 1 | 0 | 0 | 0 | 0 |
| Sandy_9   | 0 | 1 | 0 | 0 | 0 | 0 | 1 | 1 |   | 0 | 0 | 0 | 0 |
| Jerry_10  | 1 | 0 | 0 | 0 | 1 | 0 | 0 | 0 |   | 0 | 0 | 0 | 0 |
| Darrin_11 | 1 | 1 | 0 | 0 | 0 | 0 | 0 | 0 | 0 | 1 |   | 0 | 0 |
| Ben_12    | 1 | 0 | 0 | 0 | 0 | 1 | 0 | 0 | 0 | 1 | 0 |   | 0 |
| Arnie_13  | 0 | 0 | 0 | 0 | 1 | 1 | 0 | 0 | 0 | 0 | 0 | 0 |   |

*Table 8: The Transatlantic Industries team friendship network (Fine, 1987, p. 144; Doreian et al., 2005, pp. 8, 48)*

Replicating the direct structural equivalence blockmodeling of Doreian et al (2005, p. 196ff), i.e. allowing for complete and null blocks in arbitrary configurations, their Hamming-optimized solutions for partitions with 2 to 5 positions are given in the left half of Table 9 below. For each solution, corresponding blockmodels can be found in Figure 5. Repeating the blockmodel analysis when using the correlation-based goodness-of-fit measure, we arrive at the optimal solutions as given in the right-hand half of Table 9 below.

---
[13] Note that Jerry (10) and Arnie (13) only nominated two friends each (Fine, 1987, p. 144).



| | **Hamming-based optimal solutions** | | | **Correlation-based optimal solutions** | | |
|---|---|---|---|---|---|---|
| k | Hamming | (corr) | Blockmodel | corr | (Hamming) | Blockmodel |
| 2 | 29 | (0.4046) | A | 0.4046 | (29) | A |
| 3 | 23 | (0.5559) | B | 0.5752 | (27) | H |
|   | 23 | (0.5534) | C |        |      |   |
| 4 | 20 | (0.6191) | D | 0.6395 | (24) | I |
| 5 | 17 | (0.7080) | E | 0.7080 | (17) | E |
|   | 17 | (0.6871) | F |        |      |   |
|   | 17 | (0.6812) | G |        |      |   |

Table 9: Optimal solutions for conventional and correlational direct structural equivalence blockmodeling of Transatlantic Industries baseball team, 2 to 5 positions (consult Figure 5 below for respective blockmodel solution)

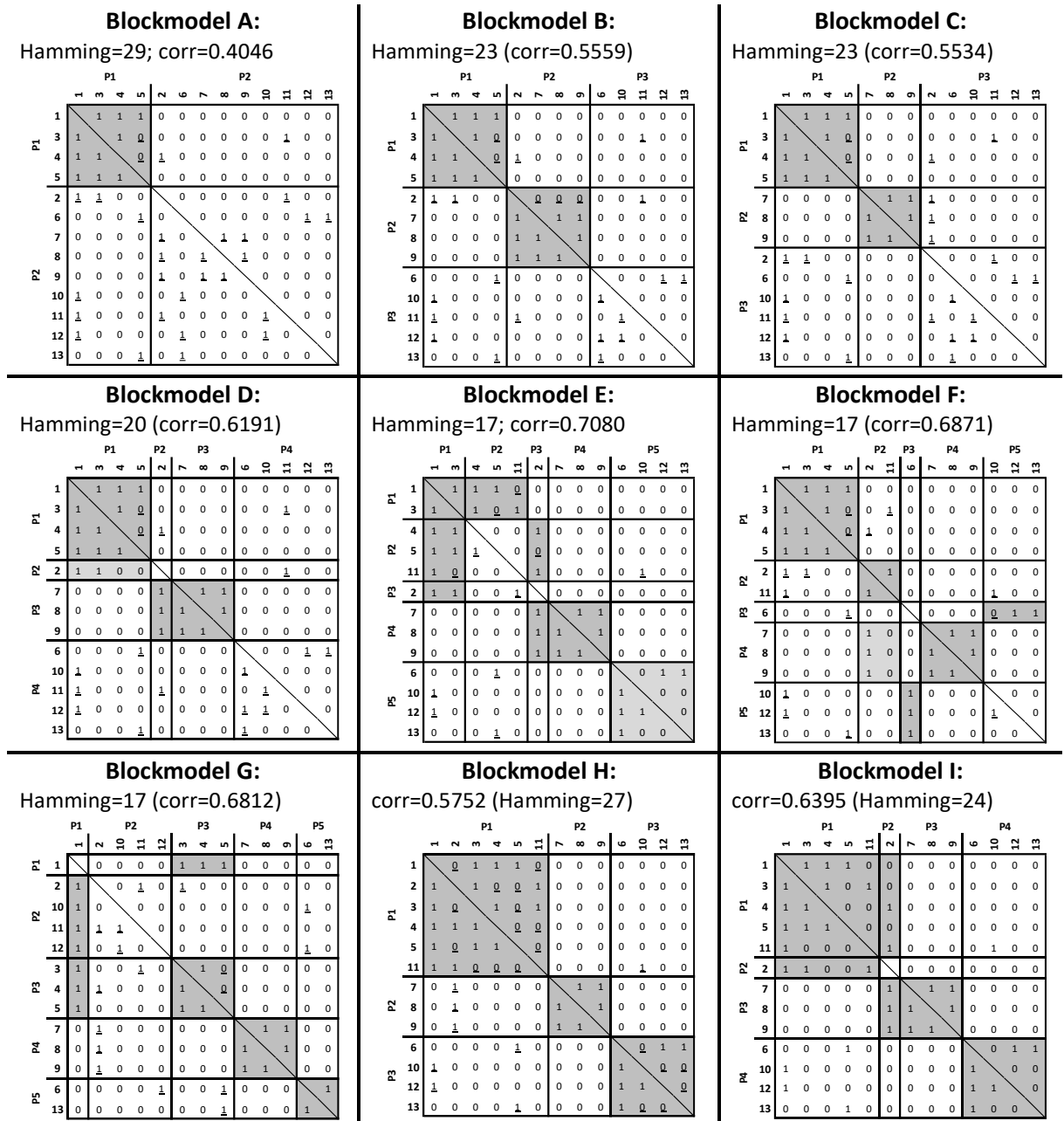

Figure 5: Optimal blockmodel solutions for the Transatlantic Industries team (dark-shaded blocks: complete blocks; light-shaded: complete or null blocks; non-shaded: null blocks; underlined cell values: block inconsistencies)



Three of the Hamming-optimized solutions in Figure 5 contain blocks that have perfect 0.5 densities. For the 4-positional solution (blockmodel D), the $B_{2,1}$ block could be either a complete or a null block, yielding the same number of inconsistencies in both cases. Similarly, one block each in two of the 5-positional solutions ($B_{5,5}$ in blockmodel E; $B_{4,2}$ in blockmodel F) have mid-point densities where either complete or null blocks would yield the same number of inconsistencies. However, the corresponding correlation-based measures differ in all three cases depending on whether we model these blocks as complete or null.

The correlation-based approach produces singular optimal blockmodel and blockimage solutions for each specified number of positions. For the 2-positional model, both heuristics arrive at the same optimal solution. When fitting to three positions, the correlation-based approach finds a singular optimal solution (blockmodel H) that has four more inconsistencies than the two optimal solutions found when using the conventional approach (blockmodel B and C). Of particular interest in the blockmodel H solution is the $B_{3,3}$ block, i.e. the intra-positional ties between John (6), Jerry (10), Ben (12), and Arnie (13). With 6 out of a total of 12 potential ties among these four kids, it would seem reasonable that this could either be categorized as a complete or a null block. However, in the correlation-based approach, this is not the case. When modeling $B_{3,3}$ as a complete block, the goodness-of-fit measure is about 10 percent higher (0.5752) than what would be the case when modeled as a null block (0.5173). These are effects from the inherent 50/50 bias of the point-biserial correlation previously discussed.

The particularities with the correlation-based heuristic is further emphasized when examining the correlation-based optimal solution for the 4-positional blockmodel (blockmodel I in Figure 5). Examining block $B_{1,2}$, we note that two individuals nominate Tom as a friend, the latter constituting a singleton position ($P_2$) in this solution. Despite this block only having 2 out of a total of 5 possible ties, the correlation-based heuristic nevertheless categorizes this as a complete block. Prespecifying with the partition and blockimage for this optimal solution, setting this block as a null block instead ($I_{1,2}$=nul), the correlation for this solution would be the slightly lower (0.6309).[14]

| k | Penalty | Correlation | Blockmodel | P1 | P2 |
|---|---|---|---|---|---|
| 2 | 3 | 0.9598 | [com,nul],[nul,reg] | 7,8,9 | rest |
|   | -"- | 0.9250 | [-,nul],[nul,reg] | 13 | rest |
|   | -"- | 0.9120 | [nul,reg],[reg,reg] | 1,2,7,12,13 | rest |
|   | -"- | -"- | -"- | 1,2,8,12,13 | rest |
|   | -"- | -"- | -"- | 1,2,9,12,13 | rest |
|   | -"- | -"- | -"- | 1,3,7,12,13 | rest |
|   | -"- | -"- | -"- | 1,3,8,12,13 | rest |
|   | -"- | -"- | -"- | 1,3,9,12,13 | rest |

*Table 10: Optimal 2-positional solutions for conventional and correlational generalized equivalence blockmodeling of Transatlantic Industries baseball team (complete, regular and null blocks in this order of priority)*

Expanding the analysis towards the generalized framework, allowing for complete, regular and null blocks (in this order), optimal solutions for the two approaches are given in Table 10 below. As

---

[14] Calculating the correlations for these two solutions using the notation for the point-biserial formula (see Eq. 1) is as follows. Both solutions have 156 datapoints ($n$) with the standard deviation $s$ of the $X$ values being ~0.4254. For the optimal solution (i.e. with $B_{1,2}$=com), $r_{pb} = \left((0.6275 - 0.0476)/s\right) \cdot \sqrt{51 \cdot 105/n^2} = 1.3632 \cdot 0.4691 = 0.6395$, and when $B_{1,2}$=nul, $r_{pb} = \left((0.6522 - 0.0636)/s\right) \cdot \sqrt{46 \cdot 110/n^2} = 1.3836 \cdot 0.4560 = 0.6309$. Thus, even though the null block increases the difference between the mean $X$ values of $M_1$ and $M_2$, this gain to the point-biserial correlation is more than offset as the ideal blockmodel (i.e. the values in $Y$) moves away from a midpoint density of 0.5.



neither goodness-of-fit measure improve when running 3-positional optimizations, only optimal 2-positional solutions are presented in this table. Where the conventional direct approach finds 8 equally optimal solutions, the correlation-based approach identifies a singular optimal solution, corresponding to the first of the 8 solutions in Table 10.

### Binary example 3: Kansas Search and Rescue communication

The final binary example consists of the communication network between organizations and units involved in a search and rescue operation following a boat accident on Lake Pomona in Kansas, USA. Collected and analyzed by Drabek et al (1981), the original data was directional and valued, capturing frequencies of communication between 20 organizations and offices involved in the emergency operation. Prior to their blockmodel analysis, Drabek and colleagues binarized this data by coding all reported communication ties as binary ties, this being the data that was subsequently analyzed by Doreian et al (2005), as well as being the data analyzed here.

In the original study by Drabek and colleagues (1981), the CONCOR algorithm (see, e.g., Wasserman & Faust, 1994, p. 376) was used to partition the network into five positions, resulting in the blockmodel and blockimage as shown in Figure 6 below.

|   |   | P1 | | P2 | | | | | P3 | | | P4 | | | | P5 | | | | | |
|---|---|---|---|---|---|---|---|---|---|---|---|---|---|---|---|---|---|---|---|---|---|
|   |   | A | E | C | F | G | I | K | D | L | N | B | H | J | S | M | O | P | Q | R | T |
| P1 | Sheriff (A) |  | 1 | 1 | 1 | 1 | 1 | 1 | 0 | 0 | 0 | 1 | 0 | 0 | 0 | 0 | 1 | 1 | 1 | 0 | 1 |
|    | HighwayP (E) | 1 |  | 1 | 1 | 1 | 0 | 1 | 1 | 1 | 1 | 0 | 0 | 0 | 0 | 1 | 0 | 1 | 1 | 0 | 1 |
| P2 | Coroner (C) | 1 | 1 |  | 1 | 0 | 1 | 1 | 1 | 0 | 0 | 0 | 0 | 0 | 0 | 0 | 0 | 0 | 1 | 0 | 0 |
|    | ParksRes (F) | 1 | 1 | 1 |  | 1 | 1 | 1 | 1 | 1 | 1 | 0 | 0 | 0 | 0 | 0 | 0 | 0 | 0 | 0 | 0 |
|    | GameFish (G) | 1 | 1 | 1 | 1 |  | 1 | 0 | 1 | 0 | 0 | 0 | 0 | 1 | 0 | 0 | 0 | 1 | 0 | 0 | 0 |
|    | ArmyCorps (I) | 1 | 1 | 1 | 1 | 1 |  | 1 | 0 | 0 | 0 | 0 | 0 | 0 | 0 | 0 | 0 | 0 | 1 | 0 | 0 |
|    | CrableAmb (K) | 1 | 1 | 1 | 1 | 1 | 1 |  | 0 | 1 | 0 | 0 | 0 | 0 | 0 | 0 | 0 | 0 | 0 | 0 | 0 |
| P3 | Attorney (D) | 1 | 1 | 1 | 1 | 1 | 1 | 0 |  | 0 | 1 | 1 | 0 | 0 | 0 | 0 | 0 | 0 | 0 | 0 | 0 |
|    | FrankCoAmb (L) | 1 | 1 | 0 | 0 | 1 | 0 | 0 | 0 |  | 0 | 0 | 0 | 0 | 0 | 0 | 0 | 0 | 1 | 0 | 0 |
|    | Shawney (N) | 1 | 1 | 1 | 1 | 1 | 1 | 0 | 0 | 1 |  | 0 | 0 | 0 | 0 | 1 | 0 | 0 | 1 | 1 | 1 |
| P4 | CivilDef (B) | 1 | 1 | 1 | 1 | 1 | 1 | 0 | 0 | 0 | 1 |  | 0 | 0 | 0 | 0 | 0 | 1 | 1 | 0 | 0 |
|    | KansasDOT (H) | 1 | 1 | 0 | 0 | 0 | 0 | 0 | 0 | 0 | 0 | 1 |  | 1 | 0 | 0 | 0 | 0 | 0 | 0 | 0 |
|    | ArmyRes (J) | 1 | 1 | 0 | 0 | 0 | 0 | 0 | 0 | 0 | 0 | 0 | 0 |  | 0 | 0 | 0 | 0 | 0 | 0 | 0 |
|    | CarbDF (S) | 1 | 1 | 0 | 1 | 1 | 1 | 0 | 0 | 0 | 0 | 1 | 0 | 0 |  | 0 | 0 | 0 | 1 | 0 | 0 |
| P5 | LeeRescue (M) | 1 | 1 | 0 | 0 | 0 | 0 | 0 | 0 | 0 | 0 | 0 | 0 | 0 | 0 |  | 0 | 0 | 0 | 0 | 0 |
|    | BurlPolice (O) | 1 | 1 | 1 | 1 | 1 | 1 | 1 | 1 | 1 | 0 | 0 | 1 | 0 | 1 |  | 1 | 1 | 1 | 1 | 1 |
|    | LyndPolice (P) | 1 | 1 | 0 | 1 | 1 | 0 | 1 | 0 | 0 | 0 | 1 | 0 | 0 | 0 | 0 | 0 |  | 0 | 0 | 0 |
|    | RedCross (Q) | 1 | 1 | 1 | 1 | 1 | 1 | 0 | 0 | 0 | 1 | 1 | 0 | 1 | 0 | 0 | 0 |  | 0 | 0 | 0 |
|    | TopicaFD (R) | 1 | 1 | 0 | 0 | 0 | 0 | 0 | 0 | 0 | 1 | 0 | 0 | 0 | 1 | 1 | 0 | 0 | 0 |  | 0 |
|    | TopekaRBW (T) | 1 | 1 | 0 | 0 | 0 | 0 | 0 | 0 | 0 | 0 | 0 | 0 | 0 | 0 | 0 | 1 | 1 | 1 | 0 |  |

*Figure 6: The 5-positional CONCOR-derived blockmodel and blockimage as found by Drabek et al (1981). Notation as in Figure 5.*

As noted by Doreian et al (2005, p. 203), the CONCOR-derived blockmodel contains 79 inconsistencies, which Doreian and colleagues deemed as less than optimal. When instead conducting a direct structural equivalence blockmodeling with five positions, Doreian et al found four equally optimal solutions, each with 57 inconsistencies (Doreian et al., 2005, pp. 205, 208). These four optimal partitions and, especially, their corresponding blockimages are indeed quite similar to each other – see Table 11 below – as such also being substantially different from the CONCOR-derived blockmodel in Drabek et al (1981).



|    | P1  | P2    | P3  | P4  | P5  | Common members | Solution-specific members | | | |
|----|-----|-------|-----|-----|-----|----------------|---|---|---|---|
|    |     |       |     |     |     |                | 1 | 2 | 3 | 4* |
| P1 | com | com   | com | nul | nul | A,E            |     |     |       |    |
| P2 | com | nul   | com | nul | nul | B,D,K,P        | +N,Q | +N,S | +N,Q,S | +S |
| P3 | com | nul   | com | nul | nul | C,F,G,I        |     |     |       | +Q |
| P4 | com | nul   | nul | nul | nul | H,J,L,M,R,T    | +S  | +Q  |       |    |
| P5 | com | com*  | com | com | nul | O              |     |     |       | +N |

*Table 11: Optimal solutions for Hamming-based direct structural equivalence (*: For solution 4, i.e. when the Shawney County Underwater Rescue Team (N) joins the Burlingame Police (O) in position 5, the ideal block $B_{5,2}$ turns into a null block; see also Doreian et al., 2005, p. 208)*

The correlation-based goodness-of-fit measure for the CONCOR-derived solution in Figure 6 above is relatively low (0.5608). For the four optimal solutions in Table 11, the corresponding correlations are all higher and almost identical, with the first solution having a slightly higher correlation ($corr_1$=0.6856) than the remaining three ($corr_3$=0.6831; $corr_4$=0.6813; $corr_2$=0.6811).

Prespecifying with the two almost identical blockimages in Table 11, i.e. with $B_{5,2}$ either being a complete or a null block, the correlation-based direct heuristic finds two of the same solutions as found when using the conventional Hamming-based approach. For the blockimage when $B_{5,2}$ is a complete block, the correlation-based approach finds the same optimal solution as solution 1 in Table 11, and solution 4 is found when $B_{5,2}$ is a null block. Thus, if the criteria function is given some slack, e.g. by also categorizing solutions with near-optimal[15] correlations as optimal, the correlation-based approach indeed seems very apt at finding the same solutions as those found when using the conventional Hamming-based criteria function.

Beyond the structural equivalence blockmodels of these previous studies, this third binary example is rounded off by fitting the Kansas Search and Rescue communication network to the ideal regular and null blocks. When fitting to a 2-positional regular blockmodel, both the conventional and weighted correlation-based direct heuristics find the same singular optimal solution (2 inconsistencies; *corr*=0.9793) as outlined in Figure 7 below (solid lines).

---

[15] Looking at the corresponding correlation-based measures for the four Hamming-optimized 57-penalty solutions in Table 11, it can be noted that the solution with the lowest correlation ($corr_2$=0.6811) is only 0.6 percent lower than the solution with the highest correlation ($corr_1$=0.6856). When finding the optimal partition when using correlations, it thus seems advisable to also explore solutions with correlation scores close to the optimal solution, either to ensure that the chosen solution sticks out from the next-best solution, or that a set of almost-equally-optimal solutions form a distinguishable cluster vis-à-vis subsequent solutions.



|  2-pos |  3-pos |  | P1 | | | | | | | | | | P1 | | | | | | | P2 | | |
|---|---|---|---|---|---|---|---|---|---|---|---|---|---|---|---|---|---|---|---|---|---|---|
| | | | P1 | | | | | | | | | | P2 | | | | | | | P3 | | |
| | | | A | D | E | F | G | I | K | N | P | Q | B | C | J | L | M | R | T | H | O | S |
| P1 | P1 | A_Sheriff |  | 1 | 1 | 1 | 1 | 1 |  | 1 |  | 1 | 1 |  |  |  |  |  | 1 | 1 |  |  |
| | | D_Attorney | 1 |  | 1 | 1 | 1 | 1 |  | 1 |  |  | 1 | 1 |  |  |  |  |  |  |  |  |  |
| | | E_HighwayP | 1 | 1 |  | 1 | 1 |  | 1 | 1 | 1 | 1 | 1 |  |  | 1 | 1 |  | 1 |  |  |  |  |
| | | F_ParksRes | 1 | 1 |  |  | 1 | 1 | 1 | 1 |  |  | 1 |  |  | 1 |  |  |  |  |  |  |  |
| | | G_GameFish | 1 | 1 | 1 | 1 |  |  |  | 1 |  |  | 1 | 1 |  |  |  |  |  |  |  |  |  |
| | | I_ArmyCorps | 1 |  | 1 | 1 | 1 |  |  | 1 |  | 1 | 1 |  |  |  |  |  |  |  |  |  |  |
| | | K_CrableAmb | 1 |  | 1 | 1 | 1 |  |  |  |  |  | 1 | 1 |  |  |  |  |  |  |  |  |  |
| | | N_Shawney | 1 |  | 1 | 1 | 1 |  |  |  |  | 1 | 1 |  |  | 1 | 1 | 1 | 1 |  |  |  |  |
| | | P_LyndPolice | 1 |  | 1 | 1 |  | 1 |  |  |  |  | 1 |  |  |  |  |  |  |  |  |  |  |
| | | Q_RedCross | 1 |  | 1 | 1 | 1 |  | 1 |  |  |  | 1 | 1 | 1 |  |  |  |  |  |  |  |  |
| P1 | P2 | B_CivilDef | 1 |  | 1 | 1 | 1 |  | 1 | 1 | 1 |  | 1 |  |  |  |  |  |  |  |  |  |  |
| | | C_Coroner | 1 | 1 | 1 | 1 |  | 1 | 1 |  |  | 1 |  |  |  |  |  |  |  |  |  |  |  |
| | | J_ArmyRes | 1 |  | 1 |  |  |  |  |  |  |  |  |  |  |  |  |  |  |  |  |  |  |
| | | L_FrankCoAmb | 1 |  | 1 |  | 1 |  |  |  |  | 1 |  |  |  |  |  |  |  |  |  |  |  |
| | | M_LeeRescue | 1 |  | 1 |  |  |  |  |  |  |  |  |  |  |  |  |  |  |  |  |  |  |
| | | R_TopicaFD | 1 |  | 1 |  |  |  |  | 1 |  |  |  |  |  |  |  | 1 |  |  |  |  |  |
| | | T_TopekaRBW | 1 |  | 1 |  |  |  |  |  | 1 | 1 |  |  |  |  |  |  |  |  |  | 1 |  |
| P2 | P3 | H_KansasDOT | 1 |  | 1 |  |  |  |  |  |  |  | 1 | 1 |  |  |  |  |  |  |  |  |  |
| | | O_BurlPolice | 1 | 1 | 1 | 1 | 1 | 1 | 1 | 1 | 1 | 1 |  | 1 | 1 | 1 | 1 | 1 | 1 |  |  |  |  |
| | | S_CarbDF | 1 |  | 1 | 1 | 1 |  |  |  |  | 1 | 1 |  |  |  |  |  |  |  |  |  |  |

**2-positional**

|  | P1 | P2 |
|---|---|---|
| P1 | reg | nul |
| P2 | reg | nul |

**3-positional**

|  | P1 | P2 | P3 |
|---|---|---|---|
| P1 | reg | reg | nul |
| P2 | reg | nul | nul |
| P3 | reg | reg | nul |

*Figure 7: Optimal regular 2- and 3-positional blockmodels for the Kansas SAR communication network (2-positional partition as solid lines; 3-positional partition as solid and dashed lines)*

Extending to 3-positional solutions, fitting the network to all unique, non-trivial regular blockimage varieties, the correlation-based heuristic finds the singular optimal blockmodel solution as outlined in Figure 7 (solid *and* dashed lines). At a goodness-of-fit correlation at 0.9727, the 3-positional solution is objectively less ideal than the optimal 2-positional solution. Despite this, the 3-positional solution does seem more intuitive, finding a singular solution that splits the previous 17-actor regular block into a tighter regular core, consisting of 10 organizations whose communications among each other form a 3-core, and a set of 7 actors that virtually only have ties to the regular core.

When instead optimizing using the conventional inconsistency-based measure, the same 3-positional solution found by the weighted correlation-approach is found (Figure 7) – as well as 2946 other solutions, all similarly having four inconsistencies each. Of these alternative solutions, it turns out that all but one are solutions for a trivial[16] extension of the 2-positional regular blockimage solution, resulting in a plethora of possible actor permutations that all fulfil the criteria for such adjacent regular blocks. Filtering out the solutions for this trivial blockimage, a single alternative 4-penalty optimized solution[17] remains: with a marginally lower correlation (0.9633) than the optimal solution (0.9727), both solutions should reasonably be viewed as equally viable, where a would-be choice between them should rest on substantive, interpretational grounds rather than a minuscule difference in a goodness-of-fit measures.

---

[16] The blockimage for these solutions is similar to the 3-positional solution in Figure 6 above, though with $B_{2,2}$ being a regular block. This implies that position 1 and 2 are virtually identical in this blockimage – the two positions in this blockimage are structurally equivalent! – leading to a large set of equivalent solutions. Notably, the weighted correlation measures are all identical for each of these permutations.
[17] *I*=[reg,reg,nul],[reg,reg,nul],[reg,nul,nul]; *P* = ( {A,B,F,G,I,J,N,Q}, {C,D,E,K,L,M,O,P,R,T}, {H,S} ).



## Valued example 1: Hlebec notesharing network

Building on survey data collected by Hlebec (1996), this network captures the self-reported sharing of notes among 13 Slovenian social-informatics students. With tie values ranging from 1 to 19 and tie directionality capturing the relation from borrower to lender, this network has been used to demonstrate the valued blockmodeling approach of Žiberna (2007a) as well as the deviational blockmodeling approach of Nordlund (2016). As the latter approach involves a transformation of the valued data, the comparisons in this section is primarily wth respect to the findings of Žiberna.

In his analysis, Žiberna (2007a) searches for optimal 3-positional blockmodels containing either conventional null blocks or sum-regular blocks, the latter being one of the novel valued block types suggested in his article. The criteria for a sum-regular block is that the sum of each individual row and column is at least $m$, the latter being a network-wide parameter that he chooses to set to 10 in this particular analysis. This results in the single solution as given in Figure 8 below.

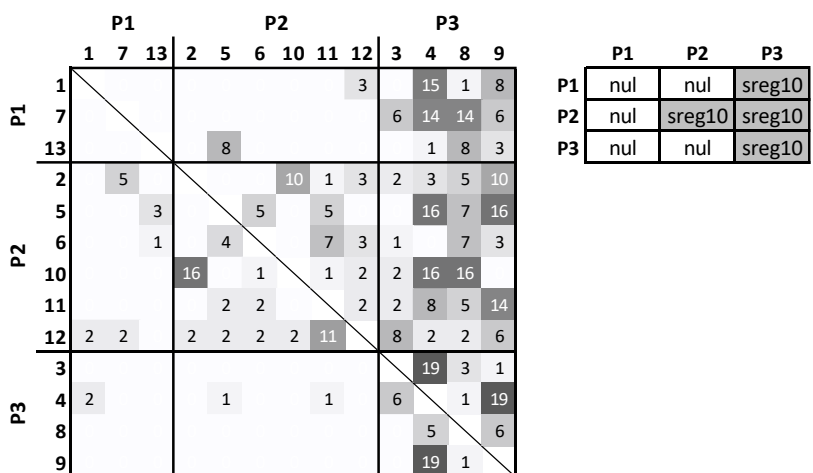

*Figure 8: Optimal blockmodel and blockimage using Žiberna's valued network approach (sreg:sum-regular blocks with m=10)*

Switching to the correlation-based heuristic, the optimal regular blockmodels and their corresponding blockimages for 2, 3 and 4 positions are given in Figure 9. The blockimage for the correlation-based 3-positional solution is identical to that found by Žiberna (see Figure 8), but their respective partitions are somewhat different, with only 8 out of 13 actors placed in the corresponding positions for respective solution. Despite this, their corresponding correlation-based measures are nevertheless quite similar: for the solution found by Žiberna using his suggested approach (Figure 8), the corresponding correlation coefficient is 0.8189, i.e. only 7 percent lower than that found when optimizing on this correlation measure (blockmodel B in Figure 9 below).

Extending to four positions, once again testing for ideal regular and null blocks, we arrive at an optimal solution with a fractionally higher correlation (0.8967) – see blockmodel C in Figure 9 below. This 4-positional optimal solution is a single-split descendant from the 3-positional optimal solution, with students 6, 12 and 13 in position $P_1$ of the 3-positional solution (blockmodel B) forming their own position in the 4-positional solution. A particular feature of these three students (i.e. $P_2$ in blockmodel C) is that they are relatively disconnected from the overall note-borrowing circus, both as lenders and borrowers, a social role not readily observed in the optimal blockimages for the optimal 2- and 3-positional solutions. Thus, even though the optimal correlation for the 4-positional solution is only marginally higher than that of the optimal 3-positional solution, the former solution might still make more sense in this particular context, i.e. that there indeed might be students that are less keen to both borrow and lend out notes to their peers.



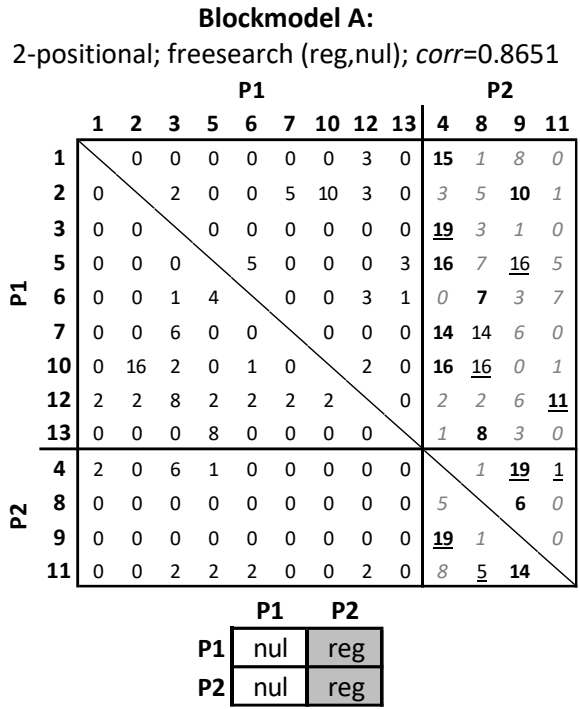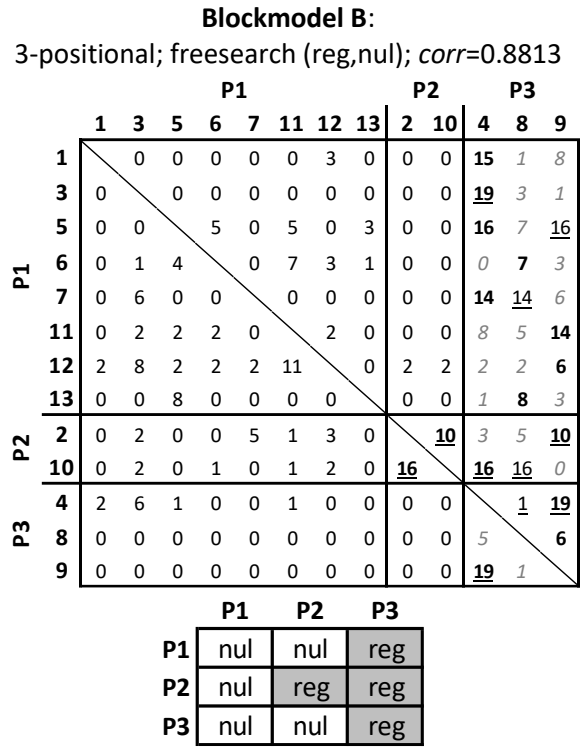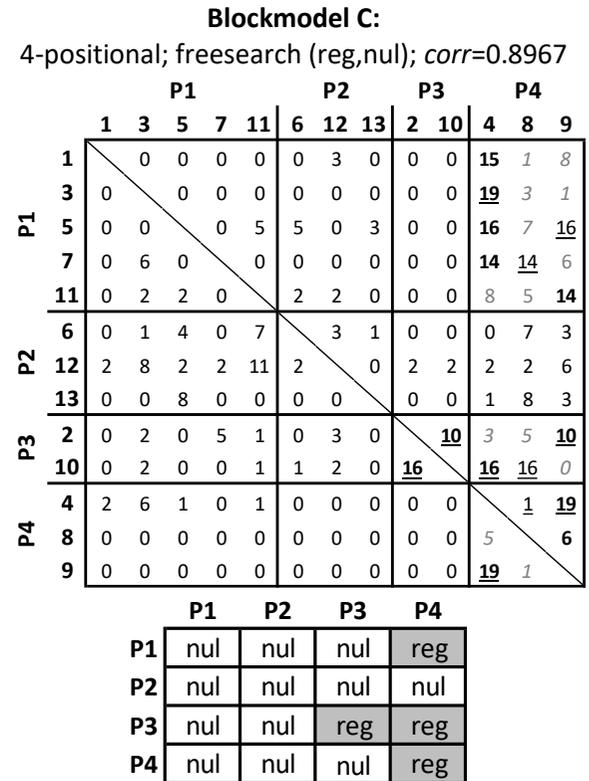

*Figure 9: Correlation-optimal regular blockmodels (k=2,3,4) of Hlebec notesharing network (cell notation for checking block criteria: bold: row-regular criteria; underline: column-regular criteria; default: null block criteria; italicized/grey: excluded)*



## Valued example 2: primates at a Florida river

Based on observations of a trope of 20 monkeys at a river in Ocala, Florida, this symmetric valued network captures the total number of joint presences at the river during a 3-month period. With 5 male and 15 female monkeys, Borgatti and Everett (2000) used this projected 2-mode network to demonstrate how their core-peirphery measure can be used with given partitions. Testing the hypothesis that core-vs-periphery membership followed a male-female division, with the off-diagonal female-male interactions modeled as complete blocks, Borgatti and Everett found a correlation of 0.2058, whose insignificance was demonstrated with a QAP-test (2000, p. 381).

Using their prespecified 2-positional core-periphery model (i.e. with the diagonal blocks modeled as complete blocks) in search of a correlation-wise optimal partition, the most optimal solution (corr=0.5463) places one male and two female monkeys in the core (blockmodel A in Figure 10 below). When allowing for any non-trivial 2-positional structural blockmodel, the optimal solution (corr=0.5821) is a core-periphery structure with null blocks in the off-diagonal blocks, with four males and eight females constituting the core – see blockmodel B in Figure 10 below.

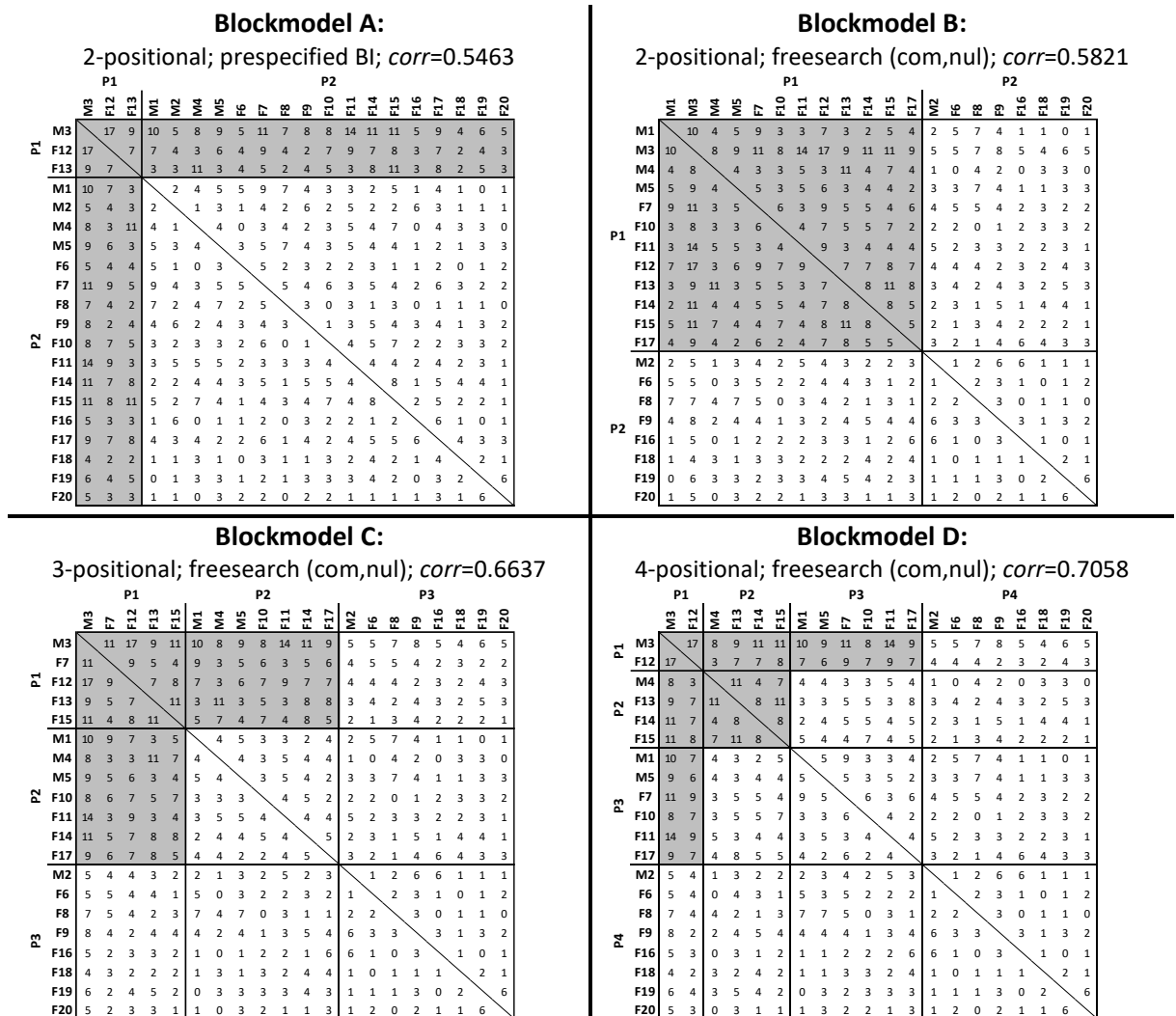

Figure 10: Correlation-optimal structural blockmodels of Florida river primates network (shaded blocks: complete blocks; non-shaded blocks: null blocks)

Extending the free-searching to three positions, a single optimal solution is found (blockmodel C in Figure 10) at a correlation of 0.6637. Both the partition and blockimage for this solution is a 'hierarchical child' of the optimal 2-positional solution. Both solutions have identical 'peripheral'



positions, both with respect to its actors and associated null blocks, and the 3-positional blockimage splits the 2-positional core to a cohesive and a non-cohesive part. Continuing to a 4-positional configuration, this pattern is repeated. At a slightly higher correlation of 0.7058, the optimal 4-positional solution (blockmodel D in Figure 10) has the same peripheral positions as in the 2- and 3-positional optimal solutions. The 4-positional blockimage extends in a similar way on the optimal 2- and 3-positional solutions. The 'peripheral' position in the 4-positional solution remains the same as in the optimal 2- and 3-positional solutions, where the remaining positions are only hierarchically linked to the grander core position of the 2-positional solution.

Switching to regular equivalence, i.e. allowing for regular and null blocks, the testing of all unique, non-trivial regular blockimages finds a singular optimal partition at a correlation of 0.8894. This optimal solution is identical to its structural-equivalence counterpart (i.e. blockmodel B in Figure 10 above) where a regular block replaces the complete block. Increasing the number of positions, the optimal 3-positional regular blockmodel has a blockimage that is also similar to its structural-equivalent counterpart (i.e. blockmodel C in Figure 10 where regular blocks replace the complete ones). As before, the disconnected 'peripheral' position in the 3-positional regular solution has the same actors as all other structural and regular optimal solutions. The optimal 3-positional regular solution does however have a slightly lower goodness-of-fit (0.8777) than its 2-positional counterpart, which together with position $P_2$ being a singleton position (holding F12) indicates that the 2-positional representation is the better choice for representing regular equivalence in this network.

Can the functional anatomy of this network be adequately expressed in terms of the more complex ideal blocks used in generalized blockmodeling? Prespecifying a core-periphery blockimage where periphery-to-core[18] relations are modeled as row-regular, the optimal solution with a correlation of 0.7556 has one male and four females in the core – see blockmodel A in Figure 11 below. Changing the ideal off-diagonal blocks to regular blocks, the same optimal partition is found, though this time at a slightly higher correlation (0.7836).

When switching the periphery-to-core block into a column-regular block, i.e. with core-to-periphery ties in the form of a row-regular block, the optimal solution – see blockmodel B in Figure 11 below – has a correlation of 0.8903. With peripheral dependency no longer being a criteria for core-periphery relations (see Nordlund, 2018, p. 354), the composition of the core is slightly different from the previous model (blockmodel A), where two female monkeys (F7 and F13) become peripheral as one additional male monkey (M4) joins the core. Of particular interest is the strong tie between M4 and F13, this being a dominance tie in both solutions in Figure 11, though where the involved monkeys swap position in the core-periphery structure that evidently seems to be there.

---

[18] And due to the symmetry of this network, core-to-periphery relations are here modeled as column-regular.



## Blockmodel A:
2-positional; prespecified BI:s (com,reg,nul,cre,rre)

|     | P1 |   |   |   |   | P2 |   |   |   |   |   |   |   |   |   |   |   |   |   |   |
|-----|----|----|----|----|----|----|----|----|----|----|----|----|----|----|----|----|----|----|----|----|
|     | M3 | F7 | F12 | F13 | F15 | M1 | M2 | M4 | M5 | F6 | F8 | F9 | F10 | F11 | F14 | F16 | F17 | F18 | F19 | F20 |
| M3  |    | 11 | 17 | 9  | 11 | 10 | 5  | 8  | 9  | 5  | 7  | 8  | 8  | 14 | 11 | 5  | 9  | 4  | 6  | 5  |
| F7  | 11 |    | 9  | 5  | 4  | 9  | 4  | 3  | 5  | 5  | 5  | 4  | 6  | 3  | 5  | 2  | 6  | 3  | 2  | 2  |
| F12 | 17 | 9  |    | 7  | 8  | 7  | 4  | 3  | 6  | 4  | 4  | 2  | 7  | 9  | 7  | 3  | 7  | 2  | 4  | 3  |
| F13 | 9  | 5  | 7  |    | 11 | 3  | 3  | 11 | 3  | 4  | 2  | 4  | 5  | 3  | 8  | 3  | 8  | 2  | 5  | 3  |
| F15 | 11 | 4  | 8  | 11 |    | 5  | 2  | 7  | 4  | 1  | 3  | 4  | 7  | 4  | 8  | 2  | 5  | 2  | 2  | 1  |
| M1  | 10 | 9  | 7  | 3  | 5  |    | 2  | 4  | 5  | 5  | 7  | 4  | 3  | 3  | 2  | 1  | 4  | 1  | 0  | 1  |
| M2  | 5  | 4  | 4  | 3  | 2  | 2  |    | 1  | 3  | 1  | 2  | 6  | 2  | 5  | 2  | 6  | 3  | 1  | 1  | 1  |
| M4  | 8  | 3  | 3  | 11 | 7  | 4  | 1  |    | 4  | 0  | 4  | 2  | 3  | 5  | 4  | 0  | 4  | 3  | 3  | 0  |
| M5  | 9  | 5  | 6  | 3  | 4  | 5  | 3  | 4  |    | 3  | 7  | 4  | 3  | 5  | 4  | 1  | 2  | 1  | 3  | 3  |
| F6  | 5  | 5  | 4  | 4  | 1  | 5  | 1  | 0  | 3  |    | 2  | 3  | 2  | 2  | 3  | 1  | 2  | 0  | 1  | 2  |
| F8  | 7  | 5  | 4  | 2  | 3  | 7  | 2  | 4  | 7  | 2  |    | 3  | 0  | 3  | 1  | 0  | 1  | 1  | 1  | 0  |
| F9  | 8  | 4  | 2  | 4  | 4  | 4  | 6  | 2  | 4  | 3  | 3  |    | 1  | 3  | 5  | 3  | 4  | 1  | 3  | 2  |
| F10 | 8  | 6  | 7  | 5  | 7  | 3  | 2  | 3  | 3  | 2  | 0  | 1  |    | 4  | 5  | 2  | 2  | 3  | 3  | 2  |
| F11 | 14 | 3  | 9  | 3  | 4  | 3  | 5  | 5  | 5  | 2  | 3  | 3  | 4  |    | 4  | 2  | 4  | 2  | 3  | 1  |
| F14 | 11 | 5  | 7  | 8  | 8  | 2  | 2  | 4  | 4  | 3  | 1  | 5  | 5  | 4  |    | 1  | 5  | 4  | 4  | 1  |
| F16 | 5  | 2  | 3  | 3  | 2  | 1  | 6  | 0  | 1  | 1  | 0  | 3  | 2  | 2  | 1  |    | 6  | 1  | 0  | 1  |
| F17 | 9  | 6  | 7  | 8  | 5  | 4  | 3  | 4  | 2  | 2  | 1  | 4  | 2  | 4  | 5  | 6  |    | 4  | 3  | 3  |
| F18 | 4  | 3  | 2  | 2  | 2  | 1  | 1  | 3  | 1  | 0  | 1  | 1  | 3  | 2  | 4  | 1  | 4  |    | 2  | 1  |
| F19 | 6  | 2  | 4  | 5  | 2  | 0  | 1  | 3  | 3  | 1  | 1  | 3  | 3  | 3  | 4  | 0  | 3  | 2  |    | 6  |
| F20 | 5  | 2  | 3  | 3  | 1  | 1  | 1  | 0  | 3  | 2  | 0  | 2  | 2  | 1  | 1  | 1  | 3  | 1  | 6  |    |

|    | P1  | P2  |
|----|-----|-----|
| P1 | com | cre |
| P2 | rre | nul |

corr=0.7556

|    | P1  | P2  |
|----|-----|-----|
| P1 | com | reg |
| P2 | reg | nul |

corr=0.7836

## Blockmodel B:
2-positional; prespecified BI:s (com, nul,rre,cre)

|     | P1 |   |   |   |   | P2 |   |   |   |   |   |   |   |   |   |   |   |   |   |   |
|-----|----|----|----|----|----|----|----|----|----|----|----|----|----|----|----|----|----|----|----|----|
|     | M3 | M4 | F12 | F15 | M1 | M2 | M5 | F6 | F7 | F8 | F9 | F10 | F11 | F13 | F14 | F16 | F17 | F18 | F19 | F20 |
| M3  |    | 8  | 17 | 11 | 10 | 5  | 9  | 5  | 11 | 7  | 8  | 8  | 14 | 9  | 11 | 5  | 9  | 4  | 6  | 5  |
| M4  | 8  |    | 3  | 7  | 4  | 1  | 4  | 0  | 3  | 4  | 2  | 3  | 5  | 11 | 4  | 0  | 4  | 3  | 3  | 0  |
| F12 | 17 | 3  |    | 8  | 7  | 4  | 6  | 4  | 9  | 4  | 2  | 7  | 9  | 7  | 7  | 3  | 7  | 2  | 4  | 3  |
| F15 | 11 | 7  | 8  |    | 5  | 2  | 4  | 1  | 4  | 3  | 4  | 7  | 4  | 11 | 8  | 2  | 5  | 2  | 2  | 1  |
| M1  | 10 | 4  | 7  | 5  |    | 2  | 5  | 5  | 9  | 7  | 4  | 3  | 3  | 3  | 2  | 1  | 4  | 1  | 0  | 1  |
| M2  | 5  | 1  | 4  | 2  | 2  |    | 3  | 1  | 4  | 2  | 6  | 2  | 5  | 3  | 2  | 6  | 3  | 1  | 1  | 1  |
| M5  | 9  | 4  | 6  | 4  | 5  | 3  |    | 3  | 5  | 7  | 4  | 3  | 5  | 3  | 4  | 1  | 2  | 1  | 3  | 3  |
| F6  | 5  | 0  | 4  | 1  | 5  | 1  | 3  |    | 5  | 2  | 3  | 2  | 2  | 4  | 3  | 1  | 2  | 0  | 1  | 2  |
| F7  | 11 | 3  | 9  | 4  | 9  | 4  | 5  | 5  |    | 5  | 4  | 6  | 3  | 5  | 5  | 2  | 6  | 3  | 2  | 2  |
| F8  | 7  | 4  | 4  | 3  | 7  | 2  | 7  | 2  | 5  |    | 3  | 0  | 3  | 2  | 1  | 0  | 1  | 1  | 1  | 0  |
| F9  | 8  | 2  | 2  | 4  | 4  | 6  | 4  | 3  | 4  | 3  |    | 1  | 3  | 4  | 5  | 3  | 4  | 1  | 3  | 2  |
| F10 | 8  | 3  | 7  | 7  | 3  | 2  | 3  | 2  | 6  | 0  | 1  |    | 4  | 5  | 5  | 2  | 2  | 3  | 3  | 2  |
| F11 | 14 | 5  | 9  | 4  | 3  | 5  | 5  | 2  | 3  | 3  | 3  | 4  |    | 3  | 4  | 2  | 4  | 2  | 3  | 1  |
| F13 | 9  | 11 | 7  | 11 | 3  | 3  | 3  | 4  | 5  | 2  | 4  | 5  | 3  |    | 8  | 3  | 8  | 2  | 5  | 3  |
| F14 | 11 | 4  | 7  | 8  | 2  | 2  | 4  | 3  | 5  | 1  | 5  | 5  | 4  | 8  |    | 1  | 5  | 4  | 4  | 1  |
| F16 | 5  | 0  | 3  | 2  | 1  | 6  | 1  | 1  | 2  | 0  | 3  | 2  | 2  | 3  | 1  |    | 6  | 1  | 0  | 1  |
| F17 | 9  | 4  | 7  | 5  | 4  | 3  | 2  | 2  | 6  | 1  | 4  | 2  | 4  | 5  | 6  | 6  |    | 4  | 3  | 3  |
| F18 | 4  | 3  | 2  | 2  | 1  | 1  | 1  | 0  | 3  | 1  | 1  | 3  | 2  | 2  | 4  | 1  | 4  |    | 2  | 1  |
| F19 | 6  | 3  | 4  | 2  | 0  | 1  | 3  | 1  | 2  | 1  | 3  | 3  | 3  | 5  | 4  | 0  | 3  | 2  |    | 6  |
| F20 | 5  | 0  | 3  | 1  | 1  | 1  | 3  | 2  | 2  | 0  | 2  | 2  | 1  | 1  | 3  | 1  | 3  | 1  | 6  |    |

|    | P1  | P2  |
|----|-----|-----|
| P1 | com | rre |
| P2 | cre | nul |

corr=0.8903

*Figure 11: Optimal 2-positional solutions, prespecified generalized equivalence ideal blocks (bold: row-regular checks; underline: column-regular checks; background-shaded: complete-block checks; default: null-block checks; gray/italics: ignored)*

### Valued example 3: Freeman's EIES friendships

The third and final valued example in this article revisits the classical Freeman's EIES acquaintance data (Freeman & Freeman, 1980). Capturing the self-reported existence and strength of friendships among 32 scholars interested in social network analysis both prior to (T1) and seven months after (T2) being introduced to a computer-based message and conferencing system, the two networks are directional and valued, with values ranging from 0 (unknown) to 4 (close personal friend). Data was also collected on the academic backgrounds of each scholar, these being either sociology (S), anthropology (A), mathematics/statistics (M), or other (O). During these seven months, 118 novel friendship ties were formed (whereas nine seemingly being lost), increasing the topological density from 66 to 77 percent and the average friendship intensity from 1.37 to 1.68.

Looking at the initially reported friendship ties at the start of the 7-month period, testing for all non-trivial structural equivalence blockimages for 2 to 5 positions, the optimal solutions for respective number of positions are given in Figure 12 below. As the number of positions increase, the optimal goodness-of-fit correlation increases as well as the detail-level of the corresponding blockimages. Whereas the 2- and 3-positional optimal solutions have mixed disciplinary memberships, the 4- and 5-positional optimal solutions reflect something akin to scholarly homophily. This is particularly visible for the group of four anthropologists – 14_A, 21_A, 43_A, and 45_A – that together with a math-statistician (40_M) and an unspecified scholar (44_O) form positions in both the 4- and 5-positional optimal solutions. Seemingly knowing each other well enough to form a complete block, they also share a tendency to *not* knowing most of the sociologists. This anthropology-dominated subset is however well connected to the primary scholar[19] in the network – 1_S – who together with another sociologist and two other anthropologists form a well-nominated core group of scholars.

---

[19] As this specific actor surely corresponds to the late professor Linton Freeman himself, the particular index number of this actor is indeed befitting.



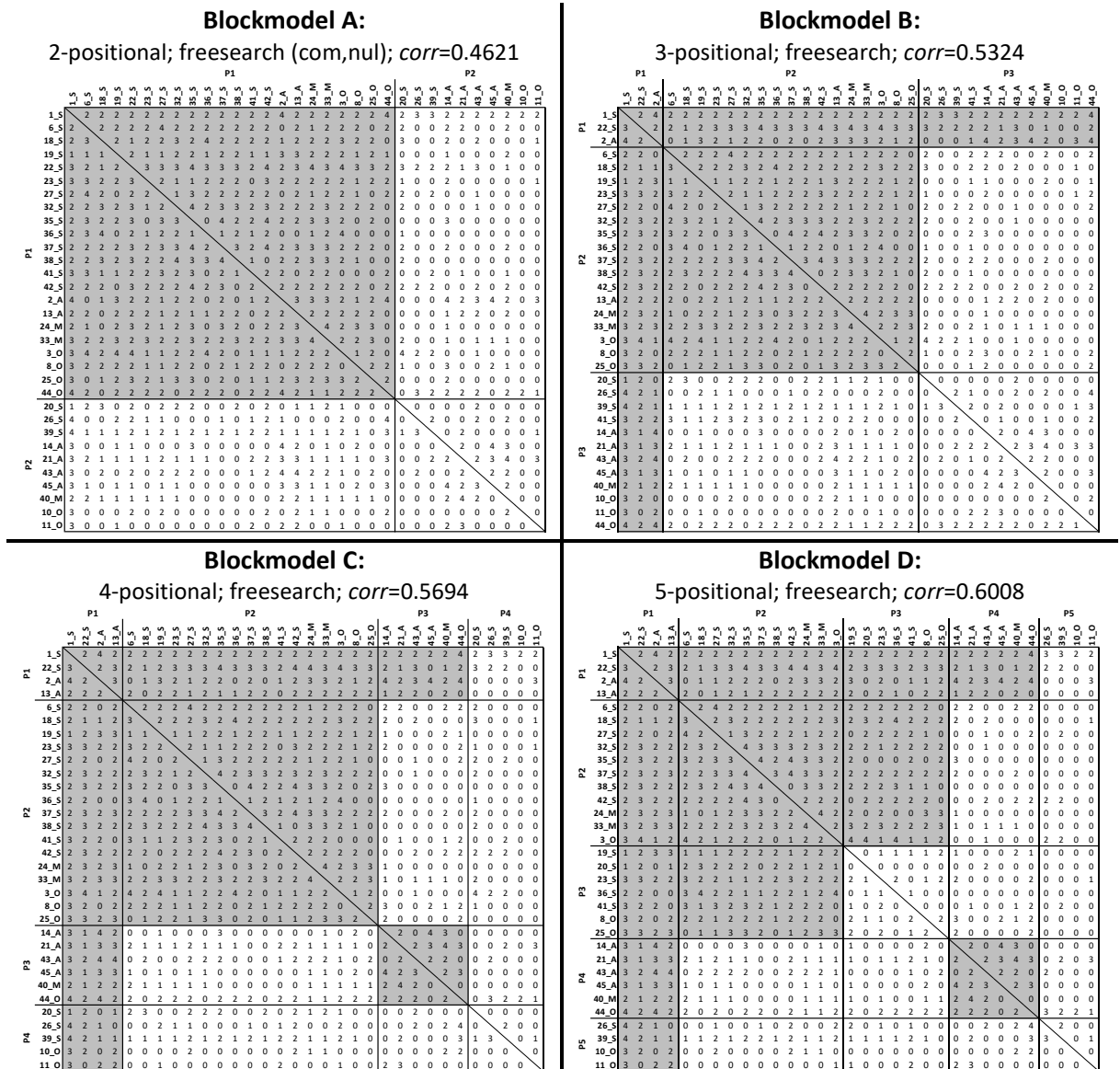

Figure 12: Correlation-optimal structural blockmodels of EIES Friendships time 1 (shaded blocks: complete blocks; non-shaded blocks: null blocks; S=sociology, A=anthropology, M=mathematics/statistics, O=other)

Instead of searching for optimal solutions, the correlation-based approach can also be used for hypothesis testing. Approaching the friendship network in this way, we can test and measure the degree of scholastic 'isolation' in the friendship networks. Prespecifying a 4-positional cohesive subgroup blockimage and partitioning the actors into discipline-specific positions, the blockmodel below (Figure 13) corresponds to both time T1 and T2. Although the correlation is low at T1 (0.2522), it is even lower at T2 (0.1770), indicating a decrease in discipline-wise isolation[20,21] over time.

---

[20] QAP-tests for both these hypothetical partitions point to significance: $p(corr_{T1} \geq 0.2522) = 0.0002$; $p(corr_{T2} \geq 0.1770) = 0.0030$.
[21] This analysis was also done with regular, rather than complete, blocks in the diagonal of the ideal blockimage, capturing a much smaller goodness-of-fit decrease from 0.7129 to 0.6920.



|    |     | Soc |     |      |     |      |      |     |      |      |     |      |     |     |     |     |     |      | Anth |     |      |     |     |     | MathSt |      |     | Other |     |      |      |      |      |
|----|-----|-----|-----|------|-----|------|------|-----|------|------|-----|------|-----|-----|-----|-----|-----|------|------|-----|------|-----|-----|-----|--------|------|-----|-------|-----|------|------|------|------|
|    |     | 1_S | 6_S | 18_S | 19_S| 20_S | 22_S |23_S | 26_S | 27_S |32_S | 35_S |36_S |37_S |38_S |39_S |41_S | 42_S | 2_A  |13_A | 14_A |21_A |43_A |45_A | 24_M   |33_M  |40_M | 3_O   | 8_O | 10_O | 11_O | 25_O | 44_O |
| Soc | 1_S |     | 2   | 2    | 2;3 | 2    | 2    | 3   | 2    | 2    | 2   | 2    | 2   | 2   | 2   | 3   | 2   | 2;3  | 4    | 2;3 | 2;3  | 2;3 | 2   | 2;3 | 2      | 2    | 2   | 2     | 2   | 2    | 2    | 2    | 4    |
|     | 6_S | 2   |     | 2;3  | 2   | 2    | 2    | 2   | 0;2  | 4    | 2   | 2    | 2   | 2   | 2   | 2   | 0;2 | 2    | 0;2  | 2   | 2    | 2;1 | 0;2 | 0   | 1;2    | 2    | 2   | 2     | 2   | 0;2  | 0    | 0    | 2    |
|     | 18_S| 2;3 | 3   |      | 2   | 3    | 1    | 2   | 0;2  | 2    | 3;4 | 2    | 4   | 2   | 2   | 0   | 2   | 2;3  | 1;0  | 2   | 2;1  | 0;2 | 2   | 0;1 | 2      | 2    | 0;2 | 3;2   | 2   | 0    | 1    | 2    | 0;2  |
|     | 19_S| 1;3 | 1;2 | 1;2  |     | 0    | 2    | 1;2 | 0;2  | 1;2  | 2   | 2    | 1   | 2   | 2   | 0   | 1;2 | 1;2  | 3;2  | 3;4 | 1;2  | 0;2 | 0   | 0;3 | 2      | 2    | 2   | 1;2   | 0;2 | 0    | 2    | 1;3  |      |
|     | 20_S| 1;2 | 2   | 3    | 0   |      | 2    | 0;1 | 0    | 2    | 2;3 | 2;1  | 0;1 | 2   | 2   | 0   | 0;1 | 2;0  | 0    | 1   | 0    | 0   | 2;1 | 0;2 | 1      | 2    | 0   | 1     | 0   | 0    | 0;1  | 0    | 0;2  |
|     | 22_S| 3   | 2;3 | 1;2  | 2   | 3    |      | 3   | 3    | 2    | 3   | 3    | 4   | 3   | 3   | 3   | 2   | 4    | 2    | 3   | 2;0  | 1;2 | 3   | 0;2 | 4      | 3    | 1;2 | 4     | 3   | 0    | 0    | 3;4  | 2;3  |
|     | 23_S| 3   | 3   | 2    | 2   | 1;2  | 3    |     | 0    | 2    | 1   | 1;2  | 2   | 2   | 2   | 0   | 0   | 3    | 2    | 2   | 2    | 0   | 0   | 0;1 | 2      | 2    | 0;1 | 2     | 1   | 0    | 1;0  | 2    | 2;3  |
|     | 26_S| 4   | 0   | 0    | 2   | 0    | 2    | 1;0 |      | 1;0  | 0   | 0    | 0   | 1;0 | 0   | 2   | 1;0 | 2    | 1;2  | 0;2 | 0    | 0;1 | 2   | 0;1 | 0      | 0    | 2   | 2     | 0   | 0    | 0;1  | 0    | 4    |
|     | 27_S| 2   | 4   | 2    | 0   | 2    | 2    | 2   | 0    |      | 1;2 | 3    | 2   | 2   | 2   | 2   | 2   | 2    | 0    | 2   | 0    | 0   | 1   | 0   | 1      | 2    | 0   | 2     | 1   | 0    | 0    | 0    | 2    |
|     | 32_S| 2   | 2   | 3    | 2   | 3    | 1    | 0   | 2    |      |     | 4    | 2   | 3   | 3   | 0   | 2   | 3    | 2    | 2   | 0    | 0   | 1   | 0   | 2      | 3    | 0   | 2     | 2   | 0    | 0    | 2    | 0    |
|     | 35_S| 2   | 3   | 2    | 2   | 0    | 3    | 0;2 | 0    | 3    | 3   |      | 0   | 4   | 2   | 0   | 2   | 2    | 2    | 3   | 2    | 0   | 0   | 0;2 | 3      | 3    | 0   | 2     | 0   | 0    | 0    | 2    | 0    |
|     | 36_S| 2;3 | 3   | 4    | 0   | 1    | 2;3  | 1   | 0    | 2;3  | 2   | 1    |     | 1;2 | 2   | 0   | 1;2 | 2;3  | 0;2  | 0;2 | 0    | 0   | 0;2 | 0   | 1      | 2;1  | 0   | 4     | 0   | 0    | 0    | 0    | 0;2  |
|     | 37_S| 2;3 | 2   | 2    | 2;3 | 2    | 3    | 2   | 0;2  | 3;2  | 3   | 4    | 2   |     | 3   | 0   | 2;3 | 4;3  | 2    | 3   | 2    | 0;2 | 0;2 | 0;2 | 3      | 3    | 2   | 2     | 2;3 | 0;2  | 0    | 2    | 0;2  |
|     | 38_S| 2   | 2;3 | 3    | 2   | 2    | 3    | 2   | 0    | 2    | 4;3 | 3    | 3   | 4;3 |     | 0   | 1   | 0;2  | 2    | 2;3 | 0    | 0   | 0   | 0   | 3      | 3    | 0   | 2     | 1   | 0    | 0    | 0    | 0    |
|     | 39_S| 4   | 1   | 1    | 1   | 1    | 2    | 1   | 3    | 2    | 1   | 2    | 1   | 2   | 1   |     | 2   | 2    | 1    | 1   | 0    | 2   | 0   | 0   | 1      | 1    | 0   | 2     | 1   | 0    | 1    | 0    | 3    |
|     | 41_S| 3   | 3   | 1    | 1   | 0    | 2    | 2   | 0    | 3    | 2   | 3    | 0   | 2   | 1   | 2   |     | 2    | 2    | 0;2 | 0    | 1;2 | 0   | 0;2 | 2      | 2    | 1   | 0     | 0   | 0    | 0    | 0    | 2    |
|     | 42_S| 2;3 | 2;3 | 2;3  | 0;2 | 2;3  | 3;4  | 2;3 | 2    | 2    | 2;3 | 4    | 2   | 3;4 | 0   | 2   | 2   |      | 2    | 2;3 | 0    | 0;2 | 2   | 0;2 | 2      | 2    | 2   | 2     | 2   | 0;2  | 0    | 0;2  | 2;3  |
| Anth| 2_A | 4   | 0;2 | 1;2  | 3   | 0    | 2    | 2   | 0;2  | 1;2  | 2   | 2    | 0;2 | 2   | 0;2 | 0;2 | 1;2 | 2    |      | 3   | 4    | 2   | 3   | 4   | 3;2    | 3;2  | 2   | 2     | 1   | 0;2  | 3;2  | 2    | 4    |
|     | 13_A| 2;3 | 2   | 0;2  | 2;4 | 0;1  | 2    | 2   | 0;2  | 1;2  | 2   | 1;2  | 1;2 | 2   | 2   | 0   | 0;1 | 2    | 2    |     | 1    | 2   | 2   | 0;4 | 2      | 2    | 2   | 2     | 0;2 | 0;1  | 2    | 0;2  |      |
|     | 14_A| 3   | 0   | 0    | 1;2 | 0    | 1    | 0   | 0    | 0    | 0   | 3;2  | 0   | 0   | 0   | 0   | 0   | 0;1  | 4    | 2   |      | 2   | 0   | 4   | 0;1    | 1;2  | 3   | 0     | 2   | 0    | 0    | 0;2  | 0;2  |
|     | 21_A| 3   | 2   | 1    | 1;2 | 0    | 1    | 1   | 0;2  | 2    | 1   | 1;2  | 1   | 0   | 0   | 2   | 2   | 2    | 3    | 3   | 2    |     | 2   | 3   | 1;2    | 1;2  | 4   | 1     | 1   | 0    | 3    | 0    | 3    |
|     | 43_A| 3   | 0;2 | 0;2  | 0   | 2    | 0    | 2   | 2    | 2    | 2   | 2    | 0;2 | 0;2 | 0   | 0   | 1;2 | 2;3  | 4;3  | 4;3 | 0    | 2   |     | 2;3 | 1      | 0    | 2   | 0     | 0   | 0    | 2    | 0;3  |      |
|     | 45_A| 3;4 | 1;2 | 0;2  | 1;3 | 0    | 1    | 0   | 0    | 1;0  | 1   | 0;3  | 0;1 | 0;1 | 0   | 1;0 | 0;2 | 0;1  | 0;3  | 3;4 | 3;4  | 4   | 2   | 3;2 |        | 1;3  | 1;2 | 0     | 2   | 0;2  | 0    | 0;2  | 3;4  |
|MathSt| 24_M| 2  | 1;2 | 0;2  | 2   | 0    | 3    | 2   | 0;2  | 2    | 3   | 0    | 3;2 | 2   | 0;2 | 1   | 2   |      | 2    | 3   | 1    | 0   | 0   | 0;2 |        | 4    | 0;1 | 2     | 3   | 0;2  | 0    | 3;2  | 0;2  |
|     | 33_M| 3   | 2   | 2    | 3   | 2    | 2    | 3   | 0    | 2    | 2   | 3    | 2   | 2   | 3   | 0   | 2   | 2    | 3    | 3   | 1    | 0   | 1   | 1   | 4      |      | 1   | 2     | 2   | 0    | 0    | 3    | 0    |
|     | 40_M| 2;3 | 2   | 1;2  | 1;2 | 0    | 1    | 1;2 | 0;2  | 1;2  | 1   | 0;2  | 0;1 | 0;2 | 0   | 0   | 0   | 0;2  | 2    | 2;3 | 2;3  | 4   | 2   | 0;2 | 1;2    | 1;2  |     | 1;2   | 1;2 | 0;2  | 0    | 1;2  | 0;2  |
|Other| 3_O | 3   | 4   | 2    | 4   | 4    | 4    | 1   | 2    | 1    | 2   | 2    | 4   | 2   | 0   | 2   | 1   | 1    | 1    | 2   | 0    | 0   | 1   | 0   | 2      | 2    | 0   |       | 1   | 0    | 0    | 2    | 0    |
|     | 8_O | 3   | 2   | 2    | 2   | 1    | 2    | 1   | 0    | 1    | 2   | 2    | 0   | 2   | 1   | 0   | 2   | 2    | 0    | 2   | 3    | 0   | 0   | 2   | 2      | 2    | 1   | 0     |     | 0    | 0    | 2    | 2    |
|     | 10_O|3;4  | 0   | 0;2  | 0;2 | 0    | 2    | 0;2 | 0    | 2    | 0   | 0    | 0   | 0;3 | 0   | 0;2 | 0;2 | 2;3  | 0;2  | 2;3 | 0    | 0   | 0   | 0;2 | 1;2    | 1;2  | 2   | 0     | 0   |      | 0    | 0    | 2;4  |
|     | 11_O| 3   | 0   | 0    | 1   | 0    | 0    | 0   | 0    | 0    | 0   | 0    | 0   | 0   | 0   | 0   | 0   | 2    | 0    | 2   | 2    | 2   | 3   | 0   | 0      | 0    | 0   | 1     | 0   | 0    |      | 0    | 0    |
|     | 25_O| 3   | 0;1 | 1    | 2;3 | 0;1  | 3    | 2   | 0;1  | 1    | 3   | 3    | 0   | 2   | 0   | 0;1 | 1   | 1;2  | 2    | 3   | 2    | 0;1 | 0;2 | 0;2 | 2;3    | 3    | 0;2 | 3     | 2   | 0    | 0    |      | 2    |
|     | 44_O| 4   | 2   | 0;2  | 2;3 | 0;2  | 2    | 2;3 | 3    | 2    | 0;2 | 2    | 2   | 2   | 2   | 0;2 | 2   | 2;3  | 4    | 2   | 2;3  | 2   | 0;4 |     | 1;2    | 1;2  | 2   | 2     | 2   | 2;3  | 1;2  | 2    |      |

*Figure 13: Testing discipline-wise isolation hypothesis, EIES friendships time 1 and 2 (single-valued cells: same value for T1 and T2; double-valued cells: left-hand values for T1 and right-hand values for T2)*

From the low goodness-of-fit score and from inspecting the blockmodel in Figure 13, it is evident that the conventional cohesive subgroups template is unsuitable for testing an hypothesis of discipline-specific role equivalence. For instance, the residual 'Others' category indeed is less internally cohesive than what is the case within, respectively, the anthropologists and mathematician-statisticians, whereas friendship among the sociologists is perhaps best captured by a regular block. Additionally, many of the off-diagonal blocks in Figure 13 above seem quite far away from the ideal null blocks of the classical cohesive-subgroup structural template.

Based on an ocular inspection of Figure 13, parts of the 4-positional blockimage is modified, allowing for alternative ideal blocks where such seem feasible – see Figure 14A. However, whereas the conventional inconsistency-minimizing direct heuristic allows for testing individual blocks independently of each other, i.e. by finding the particular ideal block that results in the least number of inconsistencies with respect to the corresponding empirical block, the correlation-based heuristics make such block-wise tests interdependent. That is, if the correlation-based heuristic finds that a regular block is a better representation of an empirical block than a null block, such a change has repercussions on the mean and standard deviations of the Y values, possibly altering how other empirical blocks are evaluated. This implies that the correlation-based heuristic is inherently global, implying that all potential blockimage configurations, such as the 384 alternative configurations in Figure 14A, must be evaluated separately.[22]

---

[22] In addition to finding optimal partitions for given blockimage configurations, the accompanying software also allows for the creation and comparative testing of such full sets of blockimage ensembles, either for given hypothetical partitions or while also finding the optimal partitions for each permuted blockimage.



**A:**

|       | Soc         | Anth    | MathSt  | Other   |
|-------|-------------|---------|---------|---------|
| Soc   | com,reg     | nul,cre | reg,rre | nul,rre |
| Anth  | nul,rre     | com     | cre     | nul,rre |
| MathSt| nul,rre,cre | nul     | com     | nul     |
| Other | reg         | reg     | nul     | reg,nul |

**B:**

|       | Soc | Anth | MathSt | Other |
|-------|-----|------|--------|-------|
| Soc   | reg | nul  | reg    | nul   |
| Anth  | nul | com  | cre    | nul   |
| MathSt| rre | nul  | com    | nul   |
| Other | reg | reg  | nul    | nul   |

*Figure 14: Alternative blockimages for discipline-specific hypothetical partition, EIES friendship time 1 (A), and the optimal blockimage found for this network and partition (B)*

From the friendship data at time 1, testing the discipline-wise 4-positional partition (see Figure 13) on the 384 alternative blockimages, the most optimal blockimage, at a correlation of 0.7232, is the one given in Figure 14B. The full blockmodel for this hypothetical partition is given in Figure 15 below. For the friendship network at time 2, an almost identical blockimage is found at a slightly lower correlation (0.7076), where the friendship nominations from the anthropologists to the sociologists is better depicted as row-regular at time point 2.

|       |      | Soc |     |      |      |      |      |      |      |      |      |      |      |      |      |      |      |      | Anth |      |      |      |      |      | MathSt |      |      | Other |     |      |      |      |      |
|-------|------|-----|-----|------|------|------|------|------|------|------|------|------|------|------|------|------|------|------|------|------|------|------|------|------|--------|------|------|-------|-----|------|------|------|------|
|       |      | 1_S | 6_S | 18_S | 19_S | 20_S | 22_S | 23_S | 26_S | 27_S | 32_S | 35_S | 36_S | 37_S | 38_S | 39_S | 41_S | 42_S | 2_A  | 13_A | 14_A | 21_A | 43_A | 45_A | 24_M   | 33_M | 40_M | 3_O   | 8_O | 10_O | 11_O | 25_O | 44_O |
| Soc   | 1_S  |     | *2* | *2* | **2** | *2* | *2* | *2* | **3** | *2* | *2* | *2* | *2* | *2* | **3** | *2* | *2* | *2* | 4 | 2 | 2 | 2 | 2 | 2 | **2** | *2* | *2* | 2 | 2 | 2 | 2 | 2 | 4 |
|       | 6_S  | 2 |    | *2* | *2* | *2* | *2* | *2* | 0 | **4** | *2* | *2* | *2* | *2* | *2* | 0 | *2* | *2* | 0 | 2 | 2 | 2 | 0 | 0 | *1* | **2** | *2* | 2 | 2 | 0 | 0 | 0 | 2 |
|       | 18_S | *2* | 3 |    | *2* | 3 | 1 | *2* | 0 | *2* | 3 | *2* | **4** | *2* | *2* | 0 | *2* | *2* | 1 | 2 | 2 | 0 | 2 | 0 | **2** | *2* | *0* | 3 | 2 | 0 | 1 | 2 | 0 |
|       | 19_S | *2* | *1* | *1* |    | 0 | **2** | *1* | 0 | *1* | *2* | *2* | *1* | *2* | *2* | 0 | *1* | *1* | 3 | 3 | 1 | 0 | 0 | 0 | **2** | *2* | *2* | 2 | 1 | 0 | 0 | 2 | 1 |
|       | 20_S | *1* | *2* | **3** | *0* |   | *2* | 0 | 0 | *2* | *2* | *2* | 0 | 0 | *2* | 0 | 0 | *2* | 0 | 1 | 0 | 0 | 2 | 0 | *1* | **2** | *0* | 1 | 0 | 0 | 0 | 0 | 0 |
|       | 22_S | *3* | *2* | *1* | *2* | 3 |    | **3** | *2* | 3 | 3 | **4** | 3 | 3 | **3** | *2* | *2* | **4** | 2 | 3 | 2 | 1 | 3 | 0 | **4** | **3** | *1* | 4 | 3 | 0 | 0 | 3 | 2 |
|       | 23_S | **3** | *3* | *2* | *2* | 1 | **3** |    | 0 | *2* | *1* | *1* | *2* | *2* | *2* | 0 | 0 | *3* | 2 | 2 | 2 | 0 | 0 | 0 | **2** | *2* | *0* | 2 | 1 | 0 | 1 | 2 | 2 |
|       | 26_S | **4** | *0* | *0* | *2* | 0 | *2* | *1* |    | *1* | 0 | 0 | *1* | *0* | *2* | *1* | *2* | *2* | 1 | 0 | 0 | 0 | 2 | 0 | *0* | *0* | **2** | 2 | 0 | 0 | 0 | 0 | 4 |
|       | 27_S | *2* | **4** | *2* | *0* | 2 | *2* | *2* | 0 |   | *1* | 3 | *2* | *2* | *2* | *2* | *2* | *2* | 0 | 2 | 0 | 0 | 1 | 0 | *1* | **2** | *0* | 2 | 1 | 0 | 0 | 0 | 2 |
|       | 32_S | *2* | *2* | *3* | *2* | 2 | *3* | *1* | 0 | *2* |   | **4** | *2* | 3 | 3 | 0 | *2* | *3* | 2 | 2 | 0 | 0 | 1 | 0 | *2* | **3** | *0* | 2 | 2 | 0 | 0 | 2 | 0 |
|       | 35_S | *2* | *3* | *2* | *2* | 0 | *3* | 0 | 0 | *3* | 3 |    | 0 | **4** | *2* | 0 | *2* | **4** | 2 | 2 | 3 | 0 | 0 | 0 | **3** | *3* | *0* | 2 | 0 | 0 | 0 | 2 | 0 |
|       | 36_S | *2* | *3* | **4** | *0* | 1 | *2* | *1* | 0 | *2* | *2* | *1* |    | *1* | *2* | 0 | *1* | *2* | 0 | 0 | 0 | 0 | 0 | 0 | *1* | **2** | *0* | 4 | 0 | 0 | 0 | 0 | 0 |
|       | 37_S | *2* | *2* | *2* | *2* | 2 | *3* | *2* | 0 | *3* | 3 | **4** | *2* |   | 3 | 0 | *2* | **4** | 2 | 3 | 2 | 0 | 0 | 0 | **3** | *3* | *2* | 2 | 2 | 0 | 0 | 2 | 0 |
|       | 38_S | *2* | *2* | *3* | *2* | 2 | *3* | *2* | 0 | *2* | **4** | 3 | 3 | **4** |   | 0 | *1* | 0 | 2 | 2 | 0 | 0 | 0 | 0 | **3** | *3* | *0* | 2 | 1 | 0 | 0 | 0 | 0 |
|       | 39_S | **4** | *1* | *1* | *1* | 1 | *2* | *1* | **3** | *2* | *1* | *2* | *1* | *2* | *1* |   | *2* | *2* | 1 | 1 | 0 | 2 | 0 | 0 | *1* | *1* | *0* | 2 | 1 | 0 | 1 | 0 | 3 |
|       | 41_S | **3** | *3* | *1* | *1* | 0 | *2* | *2* | 0 | *3* | *2* | *3* | 0 | *2* | *1* | *2* |   | *2* | 2 | 0 | 0 | 1 | 0 | 0 | **2** | *2* | *1* | 0 | 0 | 0 | 0 | 0 | 2 |
|       | 42_S | *2* | *2* | *2* | *0* | 2 | *3* | *2* | *2* | *2* | *2* | **4** | *2* | 3 | 0 | *2* | *2* |   | 2 | 2 | 0 | 0 | 2 | 0 | **2** | *2* | *2* | 2 | 2 | 0 | 0 | 0 | 2 |
| Anth  | 2_A  | 4 | 0 | 1 | 3 | 0 | 2 | 2 | 0 | 1 | 2 | 2 | 0 | 2 | 0 | 0 | 1 | 2 |     | 3 | 4 | 2 | 3 | 4 | *3* | *3* | *2* | 2 | 1 | 0 | 3 | 2 | 4 |
|       | 13_A | 2 | 2 | 0 | 2 | 0 | 2 | 2 | 0 | 1 | 2 | 1 | 1 | 2 | 2 | 0 | 0 | 2 | 2 |   | 1 | 2 | 2 | 0 | *2* | *2* | *2* | 2 | 2 | 0 | 0 | 2 | 0 |
|       | 14_A | 3 | 0 | 0 | 1 | 0 | 1 | 0 | 0 | 0 | 0 | 3 | 0 | 0 | 0 | 0 | 0 | 0 | 4 | 2 |   | 2 | 0 | 4 | *0* | *1* | *3* | 0 | 2 | 0 | 0 | 0 | 0 |
|       | 21_A | 3 | 2 | 1 | 1 | 0 | 1 | 1 | 0 | 2 | 1 | 1 | 1 | 0 | 0 | 2 | 2 | 2 | 3 | 3 | 2 |   | 2 | 3 | *1* | *1* | **4** | 1 | 1 | 0 | 3 | 0 | 3 |
|       | 43_A | 3 | 0 | 2 | 0 | 0 | 2 | 0 | 2 | 2 | 2 | 2 | 0 | 0 | 0 | 0 | 1 | 2 | 4 | 4 | 0 | 2 |   | 2 | *2* | *2* | *2* | 1 | 0 | 0 | 0 | 2 | 0 |
|       | 45_A | 3 | 1 | 0 | 1 | 0 | 1 | 0 | 0 | 1 | 1 | 0 | 0 | 0 | 0 | 0 | 0 | 0 | 3 | 3 | 4 | 2 | 3 |   | *1* | *1* | *2* | 0 | 2 | 0 | 0 | 0 | 3 |
| MathSt| 24_M | *2* | *1* | *0* | *2* | 0 | **3** | *2* | 0 | *1* | *2* | 3 | 0 | 3 | *2* | 0 | 0 | *2* | 2 | 3 | 1 | 0 | 0 | 0 |     | 4 | 0 | 2 | 3 | 0 | 0 | 3 | 0 |
|       | 33_M | **3** | *2* | *2* | *3* | 2 | *2* | *3* | 0 | *2* | *2* | *3* | *2* | *2* | *3* | 0 | 0 | *2* | 3 | 3 | 1 | 0 | 1 | 1 | 4 |   | 1 | 2 | 2 | 0 | 0 | 3 | 0 |
|       | 40_M | **2** | *2* | *1* | *1* | 0 | *1* | *1* | 0 | *1* | *1* | 0 | 0 | 0 | 0 | 0 | 0 | 0 | 2 | 2 | 2 | 4 | 2 | 0 | *1* | *1* |   | 1 | 1 | 0 | 0 | 1 | 0 |
| Other | 3_O  | *3* | **4** | *2* | **4** | **4** | **4** | *1* | *2* | *1* | *2* | *2* | **4** | **2** | *0* | **2** | *1* | *1* | 1 | **2** | 0 | 0 | 1 | 0 | 2 | 2 | 0 |   | 1 | 0 | 0 | 2 | 0 |
|       | 8_O  | **3** | *2* | *2* | *1* | 2 | *1* | 0 | *1* | *2* | *2* | 0 | *2* | **1** | *0* | **2** | *2* | *2* | 0 | **2** | **3** | 0 | 0 | 2 | 2 | 2 | 1 | 0 |   | 0 | 0 | 2 | 2 |
|       | 10_O | **3** | *0* | *0* | *0* | 2 | *0* | 0 | **2** | *0* | *0* | *0* | *0* | *0* | *0* | 0 | 0 | *2* | 0 | **2** | 0 | 0 | 0 | 0 | 1 | 1 | 2 | 0 | 0 |   | 0 | 0 | 2 |
|       | 11_O | **3** | *0* | *0* | *1* | 0 | *0* | 0 | *0* | *0* | *0* | *0* | *0* | *0* | *0* | 0 | 2 | *0* | 2 | 2 | 2 | **3** | 0 | 0 | 0 | 0 | 0 | 1 | 0 | 0 |   | 0 | 0 |
|       | 25_O | **3** | *0* | *1* | *2* | 0 | *3* | **2** | 0 | *1* | **3** | **3** | *0* | *2* | *0* | 0 | 0 | *1* | 2 | **3** | 2 | 0 | 0 | 0 | 2 | 3 | 0 | 3 | 2 | 0 | 0 |   | 2 |
|       | 44_O | **4** | *2* | *0* | *2* | 0 | *2* | *2* | **3** | *2* | *0* | *2* | *2* | *0* | *2* | *2* | *2* | *2* | **4** | 2 | 2 | 2 | **2** | 0 | 1 | 1 | 2 | 2 | 2 | 2 | 1 | 2 |   |

*Figure 15: Testing discipline-partitioned hypothesis, EIES friendships time 1 (bold: row-regular checks; underline: column-regular checks; background-shaded: complete-block checks; default: null-block checks; gray/italics: ignored)*



## Summary and conclusion

This article has proposed a correlation-based approach to direct blockmodeling for binary and valued networks. Building on the core-periphery metric and heuristic of Borgatti and Everett (2000) and its proposed extensions in Nordlund (2018), the suggested approach replaces the conventional inconsistency-based goodness-of-fit measure in direct blockmodeling with a weighted correlation coefficient. This allows for direct blockmodeling of valued as well as binary networks, using the standard set of ideal binary blocks from structural, regular and generalized blockmodeling, without any kind of dichotomization or data reduction at any point in the analysis.

Effectively bypassing the long-standing dilemma with blockmodeling of valued networks, the proposed heuristic has additional advantages. First, contrary to the conventional inconsistency-based heuristic, a standardized goodness-of-fit measure is provided, allowing for the comparison of solutions between different network sizes, value types, and equivalence notions. Additionally, benefiting from a specific particularity of point-biserial correlations, the proposed approach is inherently biased towards optimal blockmodel solutions that, it is argued, are easier to interpret. An implication of this bias is that the heuristic tends to distinguish between solutions that in the conventional approach would be deemed equally optimal, prefering solutions whose ideal blockmodels have mid-point densities of 0.5.

The proposed heuristic was demonstrated on six binary and valued networks. For the three binary examples, the proposed heuristic either finds identical or very similar solutions as those found using the conventional inconsistency-minimizing direct approach. Occasionally finding optimal solutions that are non-optimal in the conventional sense, these solutions are nevertheless argued to be more intuitive and representative than the often multiple optimal solutions found when using the conventional direct approach. The approach specified in this paper is thus not only proposed for valued networks, but also as a viable alternative, and possibly a replacement, to the conventional direct blockmodeling approach based on minimizing inconsistencies.

As demonstrated in the three valued network examples, it is demonstrated how the correlation-based approach is directly applicable to valued networks without any form of dichotomization or data reduction at any point in the analysis. For the Hlebec notesharing example, the correlation-based approach finds the identical 3-positional blockimage as that found by Ziberna using his valued blockmodeling approach. However, whereas the latter involves the introduction of a special type of valued regular block, supplemented with a specific patameter setting, the correlation-based heuristic uses the conventional ideal regular block without any parameters to set. In the second and third valued network examples, the herein proposed heuristic indeed seems apt at capturing what seems like intuitive feasible positions. In addition to using the approach for finding optimal partitions, the last EIES friendship example also demonstrates hypothesis testing and comparing goodness-of-fit measures for two separate time period.

Both the binary and valued examples illustrate how the herein proposed approach seems to distinguish between different solutions that in the conventional approach would be deemed equally optimal. However, rather than focusing solely on the particular solution with the highest correlation, it is recommended to also explore the partitions and blockmodels that are near-optimal, potentially deeming these are equally adequate representations of the underlying functional anatomy of a network. As stressed in the generalized blockmodeling aprroach, context is king: a substantive and theoretical understanding is of paramount importance in the interpretation of blockmodels (e.g. Doreian et al., 2005, p. 352).



A drawback with the herein proposed approach is that it is computationally more intensive than the conventional direct approach. Whereas local optimization searches are NP-hard, irrespective of criteria functions, the inconsistency-minimizing approach allows for the optimization of each individual block separately, independently of how other empirical blocks have been matched to ideal blocks. This is however not the case for the correlation-based approach, where the selection of a particular ideal block as representative of a particular empirical block also affects the fitting of other empirical and ideal blocks in the blockmodel. Thus, as exemplified by the 384 potential blockimages in the final Freeman EIES friendship analysis, each individual blockimage has to be tested and measured individually.

Finally, although the herein proposed approach indeed seems to work well in the examples provided in this paper, it is imperative that the approach is tested further on a much larger set of binary and valued networks. It is also imperative that the particularities of the suggested goodness-of-fit function is scrutinized further. To better understand how the weighted correlation coefficient works for the more complicated blocks of regular and generalized blockmodeling, a first step would be to derive a weighted version of the point-biserial correlation formula, disentangling the goodness-of-fit measure into the left- and right-side factors of the conventional point-biserial correlation. This and similar endeavors are however here left for future research and development.